\documentclass[a4paper,UKenglish]{lipics-v2019}

\title{On the Tractability of Un/Satisfiability} 

\author{Latif Salum}{Department of Industrial Engineering, Dokuz Eyl\"ul University, Izmir, Turkey}{latif.salum@deu.edu.tr \& latif.salum@gmail.com}{https://orcid.org/0000-0001-5660-1938}{}

\authorrunning{L. Salum}

\Copyright{Latif Salum}



\acknowledgements{I would like to thank Javier Esparza, Anuj Dawar, Avi Wigderson, Paul Spirakis, and \'{E}va Tardos, as well as anonymous reviewers for their comments and contributions throughout the development of the paper since 2008. I would like to thank Csongor Csehi from the Building Bridges II Conference. I would like to thank the faculty of the Department of Mathematics of Dokuz Eyl\"ul University, as well as my colleagues at the Industrial Engineering Department.}

\nolinenumbers 

\usepackage{tikz,algorithm,algorithmicx,algpseudocode,mathtools}

\hypersetup{
    colorlinks=true,       
    linkcolor=black!35!cyan,          
    citecolor=black!35!cyan,        
    filecolor=magenta,      
    urlcolor=magenta           
}

\newcommand{\clp}{{\searrow\hskip0.03em}}
\newcommand{\Ckk}{\mathfrak{C}}
\newcommand{\ent}{\vDash}
\newcommand{\FM}{{\varphi}}
\newcommand{\Fm}{{\phi}}
\newcommand{\GE}{\geqslant}
\newcommand{\imp}{\Rightarrow}
\newcommand{\lcl}[1]{\tilde{\scp}_{\hskip-0.07em #1}}
\newcommand{\lcln}{\tilde{\Fm}}
\newcommand{\LE}{\leqslant}
\newcommand{\Li}{\mathfrak{L}}
\newcommand{\lt}{r}
\newcommand{\nlt}{\overline{\lt}}
\newcommand{\NP}{\textsf{\bf NP}}
\newcommand{\nx}[1]{\overline{x}_{#1}}
\newcommand{\ovrl}{\tilde{\varphi}}
\newcommand{\PP}{\textsf{\bf P}}
\newcommand{\prv}{\vdash}
\newcommand{\rdc}{{\,\rightarrowtail\,}}
\newcommand{\Rdc}{{\,\rightarrow\,}}
\newcommand{\sat}{{\ent_{\alpha}}}
\newcommand{\scp}{\psi}
\newcommand{\unsat}{\nvDash}
\newcommand{\X}{\odot}

\newcommand{\Co}{(x_3 \X x_4 \X \nx5)}
\newcommand{\Ct}{(x_3 \X x_6 \X \nx7)}
\newcommand{\CT}{(x_4 \X x_6 \X \nx7)}
\newcommand{\Cf}{(x_3 \hskip0.1em\wedge\hskip0.1em \nx4 \wedge x_5)}
\newcommand{\Cfb}{(x_3 \hskip0.1em\wedge\hskip0.1em {\color{white!30!blue}\nx4} \wedge x_5)}
\newcommand{\CF}{(x_3 \hskip0.1em\wedge\hskip0.1em \nx6 \wedge x_7)}
\newcommand{\CFb}{(x_3 \hskip0.1em\wedge\hskip0.1em {\color{white!30!blue}\nx6} \wedge x_7)}
\newcommand{\Cs}{(\phantom{x_4 \X} \;\: x_6 \X \nx7)}
\newcommand{\CS}{(\phantom{x_4 \X x_6 \X} \,\,\, \nx7)}

\newcommand{\co}{(\phantom{x_3 \X} x_4 \X \nx5)}
\newcommand{\ct}{(\phantom{x_3 \X} x_6 \X \nx7)}
\newcommand{\cT}{(x_4 \X x_6 \X \nx7)}
\newcommand{\cf}{(\phantom{x_3 \X} x_4 \phantom{\X \nx5}\;\:)}
\newcommand{\cF}{(\phantom{x_3 \X x_6 \X} \nx7)}
\newcommand{\cs}{(x_4 \hskip0.06em\wedge\hskip0.06em \nx6 \hskip0.06em\wedge\hskip0.05em x_7)}

\newcommand{\Cmmnt}[1]{{\small\color{black!60!gray}\Comment{#1}}}

\begin{document}

\maketitle

\begin{abstract}
This paper shows $\PP = \NP$ via exactly-1 3SAT (X3SAT). Let $\Fm = \hskip-0.11em\bigwedge\hskip-0.11em C_{k\hskip-0.09em}$ be some X3SAT formula. $C_{k\hskip-0.14em} = (\lt_{i\hskip-0.07em} \X \lt_{j\hskip-0.07em} \X \lt_u)$ is a clause denoting an exactly-1 disjunction $\X$ of literals $\lt_{i\hskip-0.05em}$, $\lt_{i\hskip-0.12em} \in \{x_i, \nx{i}\}$. $C_{k\hskip-0.12em}$ is satisfied iff $(\lt_{i\hskip-0.05em} \wedge \nlt_{j\hskip-0.08em} \wedge \nlt_u) \vee (\nlt_{i\hskip-0.05em} \wedge \lt_{j\hskip-0.08em} \wedge \nlt_u) \vee (\nlt_{i\hskip-0.05em} \wedge \nlt_{j\hskip-0.08em} \wedge \lt_u)$ is satisfied, because any $C_{k\hskip-0.07em}$ contains \emph{exactly one} true literal by the definition of X3SAT. Let $\Fm(\lt_j) := \lt_{j\hskip-0.09em} \wedge \Fm$. Then, $\lt_{j\hskip-0.09em}$ leads to reductions due to $\X$ of some $C_{k\hskip-0.14em} = (\nx{i\hskip-0.07em} \X \lt_{j\hskip-0.09em} \X x_u)$ into $c_{k\hskip-0.14em} = x_{i\hskip-0.07em} \wedge \lt_{j\hskip-0.09em} \wedge \nx{u\hskip-0.05em}$, and some $C_{k\hskip-0.14em} = (\nlt_{j\hskip-0.09em} \X \lt_{u\hskip-0.09em} \X \lt_v)$ into $C_{k'\hskip-0.15em} = (\lt_{u\hskip-0.09em} \X \lt_v)$. As a result, $\lt_{j\hskip-0.12em}$ transforms $\Fm$ into $\Fm(\lt_j) = \scp(\lt_j) \wedge \Fm'(\lt_j)$, unless $\unsat \scp(\lt_j)$, that is, unless $\scp(\lt_j)$ involves a contradiction $x_{i\hskip-0.06em} \wedge \nx{i\hskip-0.07em}$. Also, $\scp(\lt_j)$ and $\Fm'(\lt_j)$ become \emph{disjoint}, where $\scp(\lt_j) \hskip-0.05em= \hskip-0.11em\bigwedge\hskip-0.03em (c_{k\hskip-0.05em} \wedge C_{k'\hskip-0.09em})$ for $|C_{k'}| = 1$, and $\Fm'(\lt_j) \hskip-0.03em=  \hskip-0.11em\bigwedge\hskip-0.03em (C_{k\hskip-0.05em} \wedge C_{k'\hskip-0.07em})$. It is trivial to verify $\unsat \scp(\lt_j)$ and \emph{redundant} to verify $\unsat \Fm'(\lt_j)$, thus \emph{easy} to verify $\unsat \Fm(\lt_j)$.  A proof is sketched as follows. $\Fm$ transforms into $\scp \wedge \Fm'\hskip-0.10em$ such that whenever $\unsat \scp(\lt_j)$, $\nlt_{j\hskip-0.12em}$ is placed in $\scp$, and leads to reductions of some $C_{k\hskip-0.12em}$ in $\Fm'\hskip-0.10em$. If $\scp$ involves $x_{j\hskip-0.12em} \wedge \nx{j\hskip-0.05em}$, then $\Fm$ is unsatisfiable. Otherwise, $\Fm$ is satisfiable, because $\Fm$ is composed of $\scp, \scp(\lt_{i_0}), \scp(\lt_{i_1\hskip-0.05em} | \lt_{i_0}), \ldots, \scp(\lt_{i_n} | \lt_{i_m})$, and all $\scp(.)$ are \emph{disjoint} and \emph{satisfied}. Note that $\lt_{i\hskip-0.09em} \ent \scp(\lt_i)$ and $\scp(\lt_i) \ent \scp(\lt_i | .)$ for any $\lt_{i\hskip-0.09em}$ in $\Fm'\hskip-0.10em$. Thus, $\Fm'(\lt_i)$ is \emph{satisfiable}, because $\Fm \equiv \scp(\lt_i) \wedge \Fm'(\lt_i)$, where $\scp(\lt_i)$ and $\Fm'(\lt_i)$ are \emph{disjoint}. Therefore, it is \emph{redundant} to check if $\unsat \Fm'(\lt_i)$ to verify $\unsat \Fm(\lt_i)$, QED. The time complexity is $O(mn^3)$. Therefore, $\PP = \NP$.
\end{abstract}

\section{\texorpdfstring{Introduction: Effectiveness of X3SAT in proving $\PP = \NP$}{Introduction}}
As is well known, $\PP = \NP$, if there exists an efficient algorithm for any \emph{one} of $\NP$-complete problems. That is, their algorithmic efficiency is \emph{equivalent}. Nevertheless, some $\NP$-complete problem features algorithmic effectiveness, if it incorporates an \emph{effective} tool to develop an efficient algorithm. That is, a particular problem can be more effective to prove $\PP = \NP$. This issue might also be related to ``complexity reductions'' (Lipton and Regan \cite{GdlLostLttr}). They state these reductions are needed to understand what the $\PP = \NP$ problem is really about.

The paper shows that one-in-three SAT, which is $\NP$-complete~\cite{Sch78}, features algorithmic effectiveness to prove $\PP = \NP$. This problem is also known as exactly-1 3SAT (X3SAT). It incorporates ``exactly-1 disjunction'', denoted by $\X$, the tool used to develop an efficient (or a polynomial time) algorithm, which ``scans'' an X3SAT formula $\Fm$, thus is called the $\Fm$ scan.

If $\unsat \Fm(\lt_j)$, that is, $\Fm(\lt_j)$ is unsatisfiable, then $\lt_{j\hskip-0.12em}$ is incompatible, where $\Fm(\lt_j) \hskip-0.07em\coloneqq \lt_{j\hskip-0.09em} \wedge \Fm$ and $\lt_{j\hskip-0.12em} \in \{x_j, \nx{j}\}$. The $\Fm$ scan removes each incompatible $\lt_{j\hskip-0.12em}$ from $\Fm$, thus verifies compatibility of any $\lt_{i\hskip-0.09em}$ for satisfying $\Fm$. When each $\lt_{j\hskip-0.12em}$ incompatible is removed, $\Fm$ is unsatisfiable, or satisfiable. If $\Fm$ is satisfiable, then any $\lt_{i\hskip-0.12em}$ becomes compatible to participate in a satisfying assignment.

Let $\Fm = C_{1\hskip-0.12em} \wedge C_{2\hskip-0.07em} \wedge \cdots \wedge C_{m\hskip-0.12em}$ be an X3SAT formula, in which a clause $C_{k\!} = (\lt_{i\hskip-0.09em} \X \lt_{j\hskip-0.09em} \X \lt_u)$ is an exactly-1 disjunction of literals. $C_{k\hskip-0.12em}$ is satisfied by definition iff \emph{exactly one} of $\lt_{i\hskip-0.05em}$, $\lt_{j\hskip-0.05em}$, or $\lt_{u\hskip-0.09em}$ is true. Note that $(\lt_i \hskip-0.09em \vee \hskip-0.09em \lt_j \hskip-0.09em \vee \hskip-0.09em \lt_u)$ in a 3SAT formula is satisfied iff at least one of them is true.

Incompatibility of $\lt_{i\hskip-0.12em}$ is checked by a \emph{deterministic} chain of \emph{reductions} of some $C_{k\hskip-0.09em}$ in $\Fm(\lt_i)$. Consider $\Fm(x_j) \hskip-0.07em\coloneqq x_{j\hskip-0.09em} \wedge \Fm$. Then, the reductions are initiated by $x_{j\hskip-0.05em}$, and followed by $\neg \nx{j\hskip-0.05em}$, since $x_{j\hskip-0.09em} \imp \neg \nx{j\hskip-0.05em}$. That is, each $(x_{j\hskip-0.12em} \X \nx{i\hskip-0.07em} \X x_u)$ \emph{collapses} to $(x_{j\hskip-0.12em} \wedge x_{i\hskip-0.09em} \wedge \nx u)$ due to $x_{j\hskip-0.09em} \imp x_{j\hskip-0.12em} \wedge \neg \nx{i\hskip-0.09em} \wedge \neg x_{u\hskip-0.05em}$, since there is exactly one (negated) variable that is true in any $C_{k\hskip-0.09em}$ by the definition of X3SAT. Also, each $(\nx{j\hskip-0.12em} \X \nx{u\hskip-0.11em} \X x_{v\hskip-0.02em})$ \emph{shrinks} to $(\nx{u\hskip-0.11em} \X x_{v\hskip-0.02em})$ due to $\neg \nx{j\hskip-0.05em}$. As a result, $x_{j\hskip-0.12em}$ transforms $\Fm$ into $\Fm(x_j) = x_{j\hskip-0.09em} \wedge x_{i\hskip-0.07em} \wedge \nx{u\hskip-0.09em} \wedge \Fm^*\hskip-0.12em$, and $x_{i\hskip-0.07em} \wedge \nx{u\hskip-0.11em}$ proceeds the reductions in $\Fm^*\hskip-0.12em$, which involves $(\nx{u\hskip-0.11em} \X x_{v\hskip-0.02em})$.

The reductions over $\Fm_s(x_j)$ terminate iff $x_{j\hskip-0.12em}$ transforms $\Fm_{s\hskip-0.07em}$ into $\scp_s(x_j) \wedge \Fm'_s(x_j)$, in which $\scp_s(x_j)$ and $\Fm'_s(x_j)$ are disjoint, where $s$ denotes the current scan, and $\scp_s(x_j)$ is a conjunction of (negated) variables that are true. They are interrupted iff $\scp_s(x_j)$ involves $x_{i\hskip-0.07em} \wedge \nx{i\hskip-0.05em}$, hence $\unsat \Fm_s(x_j)$, thus $x_{j\hskip-0.12em}$ is incompatible. Note that $\unsat \Fm_s(.)$ is verified \emph{only} by $\color{red}\unsat \scp_s(.)$ (see Figure \ref{f:Assmpt}).

The reductions over $\Fm$ terminate iff $\Fm$ transforms into $\scp \wedge \Fm'\hskip-0.12em$, in which $\scp$ and $\Fm'\hskip-0.12em$ are disjoint, where $\scp = \nx{5\hskip-0.07em} \wedge x_{n\hskip-0.09em} \wedge \cdots \wedge \nx{2\hskip-0.12em}$ (Figure \ref{f:Assmpt}). Then, $\Fm$ is updated, that is, $\Fm \gets \Fm'\hskip-0.09em$. The $\Fm_{s\hskip-0.07em}$ scan is interrupted iff $\scp_{s\hskip-0.07em}$ involves $x_{i\hskip-0.09em} \wedge \nx{i\hskip-0.09em}$ for some $s$ and $i$, thus $\unsat \Fm$, that is, $\Fm$ is unsatisfiable.

\begin{figure} [!h]
\vspace{-0.58em}
\begin{center}\large
  \begin{tikzpicture}[scale=0.8, transform shape]
    \draw[|-|] +(0,0) node[left]  {$\Fm$} to +(10,0) node[right] {$\Fm_2 \hskip-0.09em\coloneqq \Fm(\nx5)$};
             \draw[|-|,red]    (1.8,0) to node[above] {{\color{black}$\neg x_{5\hskip-0.09em} \imp \nx{5\hskip-0.09em}$ for $\Fm$, if} $\unsat \scp(x_5)$}    (7,0);

    \draw[|-|] +(0,-0.8) node[left]  {$\Fm_2$} to +(10,-0.8) node[right] {$\Fm_3 \hskip-0.09em\coloneqq \Fm_2(x_n)$};
             \draw[|-|,red]    (2.2,-0.8) to node[above] {{\color{black}$\neg \nx{n\hskip-0.09em} \imp x_{n\hskip-0.09em}$ for $\Fm_2$, if}     $\unsat \scp_2(\nx n)$} (8,-0.8);

    \node at (0.3,-1.2) {$\vdots$}; \node at (9.7,-1.2) {$\vdots$};

    \draw[|-|] +(0,-1.8) node[left]  {$\Fm_{s-1}$} to +(10,-1.8) node[right] {$\Fm_s \hskip-0.09em\coloneqq \Fm_{s-1}(\nx2)$};
        \draw[|-|,red]            (3,-1.8) to node[above] {{\color{black}$\neg x_{2\hskip-0.09em} \imp \nx{2\hskip-0.05em}$, if} $\unsat \scp_{s-1}(x_2)$} (9,-1.8);
  \end{tikzpicture}
\end{center}\vspace{-1.18em}
  \caption{The $\Fm_{s\hskip-0.07em}$ scan: $\unsat \Fm_s(\lt_j)$ is verified \emph{solely} by $\unsat \scp_s(\lt_j)$\,---\,whether or not $\unsat \Fm'_s(\lt_j)$ is \emph{ignored}}\label{f:Assmpt}
\vspace{-1em}
\end{figure}

\begin{claim}\label{cl:first}
$\unsat \Fm(\lt_j)$ iff $\unsat \scp_s(\lt_j)$ for some $s$. That is, it is \emph{redundant} to check whether or not $\unsat \Fm'_s(\lt_j)$. Thus, $\Fm(\lt_i)$ reduces to $\scp(\lt_i)$ due to $\Fm(\lt_i) = \scp(\lt_i) \wedge \Fm'(\lt_i)$. Then, $\scp(\lt_i) \equiv \Fm(\lt_i)$. Therefore, $\Fm$ is satisfiable iff $\scp(\lt_i)$ is \emph{satisfied} for any $\lt_{i\hskip-0.05em}$, that is, iff the $\Fm_{s\hskip-0.07em}$ scan \emph{terminates}.
\end{claim}
\begin{claimproof}[Sketch of proof]
$\scp(\lt_i)$/$\scp(\lt_i | \lt_j)$ is constructed over $\Fm$/$\Fm'(\lt_j)$, thus $\scp(\lt_i)$ \emph{covers} $\scp(\lt_i | \lt_j)$, hence $\scp(\lt_i) \ent \scp(\lt_i | \lt_j)$ holds. Because $\scp(\lt_j)$ and $\Fm'(\lt_j)$ are disjoint, $\scp(\lt_j)$ and $\scp(\lt_i | \lt_j)$ are disjoint (see Figure \ref{f:OvCntn}). Therefore, $\scp(\lt_{i_0})$, $\scp(\lt_{i_1\hskip-0.05em} | \lt_{i_0})$, $\scp(\lt_{i_2} | \lt_{i_0},\lt_{i_1\hskip-0.05em})$, and $\scp(\lt_{i_3} | \lt_{i_0}, \lt_{i_1\hskip-0.07em}, \lt_{i_2})$ form \emph{disjoint} minterms $\scp(.) = \hskip-0.15em\bigwedge\hskip-0.15em \lt_{i\hskip-0.09em}$ over $\Fm$ such that $\scp(\lt_{i_0})$, $\scp(\lt_{i_1\hskip-0.05em} | \lt_{i_0})$, $\scp(\lt_{i_2} | \lt_{i_0}, \lt_{i_1\hskip-0.05em})$, and $\scp(\lt_{i_3} | \lt_{i_0}, \lt_{i_1\hskip-0.07em}, \lt_{i_2})$ hold, since $\scp(\lt_i)$ is true for any $\lt_{i\hskip-0.12em}$ (the $\Fm_{s\hskip-0.07em}$ scan terminates), and $\scp(\lt_i) \ent \scp(\lt_i | .)$ holds. Thus, $\Fm$ is composed of $\scp(.)$ that are \emph{disjoint} and \emph{satisfied} (see Figure \ref{f:OvCntnAll}), hence $\Fm$ is satisfied.
\end{claimproof}

\begin{figure} [!h]
\vspace{-1.28em}
\begin{center}\large
  \begin{tikzpicture}[scale=0.8, transform shape]
    \draw[|-|]      +(0,0) node[left]        {$\Fm$}                    to  (12,   0);
     \draw[-|]       (0,0) to node[above]    {$\scp(\lt_i) = \lt_{i\hskip-0.07em} \wedge \lt_{j\hskip-0.09em} \wedge \cdots \wedge \lt_{v\hskip-0.05em}$}  ( 5,   0);

    \draw[|-|]      +(0,-0.8) node[left]     {$\Fm(\lt_j)$}             to  (12,-0.8);
     \draw[-|]       (0,-0.8) to node[above] {$\scp(\lt_j)$}                ( 3,-0.8);  \draw (3,-0.8) to node[above] {$\Fm'(\lt_j)$} (12,-0.8);

    \draw[|-|]      +(3,-1.6) node[left]     {$\Fm'(\lt_j) \ni \lt_i$}  to  (12,-1.6);
     \draw           (3,-1.6) to node[above] {$\scp(\lt_i | \lt_j) = \lt_{i\hskip-0.07em} \wedge \cdots \wedge \lt_{v\hskip-0.05em}$}                       ( 7,-1.6);
     \draw[|-]       (7,-1.6) to node[above] {$\Fm'(\lt_i | \lt_j)$}        (12,-1.6);
  \end{tikzpicture}
\end{center}\vspace{-1.18em}
  \caption{$\scp(\lt_i) \ent \scp(\lt_i | \lt_j)$, and $\scp(\lt_j)$ and $\scp(\lt_i | \lt_j)$ are disjoint, thus $\scp(\lt_j) \wedge \scp(\lt_i | \lt_j)$ is true}\label{f:OvCntn}
\vspace{-0.38em}
\end{figure}

A satisfying assignment $\alpha$ is constructed by composing $\scp(.)$ that are \emph{disjoint} and \emph{satisfied}. For example, $\alpha = \{\scp, {\color{purple} \scp(\lt_{i_0})}, {\color{white!30!blue} \scp(\lt_{i_1\hskip-0.05em} | \lt_{i_0})}, {\color{brown} \scp(\lt_{i_2} | \lt_{i_0}, \lt_{i_1\hskip-0.05em})}, {\color{green!40!black} \scp(\lt_{i_3} | \lt_{i_0}, \lt_{i_1\hskip-0.07em}, \lt_{i_2})}\}$ (see Figure \ref{f:OvCntnAll}).

\begin{figure} [!h]
\vspace{-0.38em}
\begin{center}\large
  \begin{tikzpicture}[scale=0.8, transform shape]
  \draw[|-|, thick] +(0,0) node[left] {$\Fm$} to (12,0);
    \draw[|-|,white!30!blue] (0  ,0.4) to node[above] {$\scp(\lt_{i_1})$} (2.6,0.4); \draw[|-|,purple] (2.4,0.2) to node[above] {$\scp(\lt_{i_0\hskip-0.05em})$} (4.0,0.2);
    \draw[|-|,green!40!black]          (3.8,0.4) to node[above] {$\scp(\lt_{i_3})$} (8.5,0.4); \draw[|-|,brown]  (6.9,0.2) to node[above] {$\scp(\lt_{i_2})$}              (12 ,0.2);

  \draw[|-|,purple]         +(0  ,-1  ) node[left,black] {$\Fm(\lt_{i_0})$}                                                      to node[above]
                                                         {$\scp(\lt_{i_0})$}                                                       (1.6,-1  );
   \draw[-|]                 (1.6,-1  ) to node[above]   {$\Fm'(\lt_{i_0})$}                                                       (12 ,-1  );

  \draw[|-|,white!30!blue]  +(1.6,-1.8) node[left,black] {$\Fm'(\lt_{i_0}) \ni \lt_{i_1}$}                                       to node[above]
                                                         {$\scp(\lt_{i_1\hskip-0.05em} | \lt_{i_0})$}                              (4.1,-1.8);
   \draw[-|]                 (4.1,-1.8) to node[above]   {$\Fm'(\lt_{i_1\hskip-0.05em} | \lt_{i_0})$}                              (12 ,-1.8);

  \draw[|-|,brown]          +(4.1,-2.6) node[left,black] {$\Fm'(\lt_{i_1\hskip-0.05em} | \lt_{i_0}) \ni \lt_{i_2}$}              to node[above]
                                                         {$\scp(\lt_{i_2} | \lt_{i_0}  , \lt_{i_1\hskip-0.05em})$}                 (7.4,-2.6);
   \draw[-|]                 (7.4,-2.6) to node[above]   {$\Fm'(\lt_{i_2} | \lt_{i_0}  , \lt_{i_1\hskip-0.05em})$}                 (12 ,-2.6);

  \draw[|-|,green!40!black]           +(7.4,-3.4) node[left,black] {$\Fm'(\lt_{i_2} | \lt_{i_0}  , \lt_{i_1\hskip-0.05em}) \ni \lt_{i_3}$} to node[above]
                                                         {$\scp(\lt_{i_3} | \lt_{i_0}  , \lt_{i_1\hskip-0.05em}, \lt_{i_2})$}      (12 ,-3.4);
  \end{tikzpicture}
\end{center}\vspace{-1.18em}
  \caption{$\color{white!30!blue}\scp(\lt_{i_1\hskip-0.05em}) \ent \scp(\lt_{i_1\hskip-0.05em} | \lt_{i_0})$, $\color{brown}\scp(\lt_{i_2}) \ent \scp(\lt_{i_2} | \lt_{i_0}, \lt_{i_1\hskip-0.05em})$, and $\color{green!40!black}\scp(\lt_{i_3}) \ent \scp(\lt_{i_3} | \lt_{i_0}, \lt_{i_1\hskip-0.05em}, \lt_{i_2})$} \label{f:OvCntnAll}
\vspace{-1.5em}
\end{figure}
\section{Basic Definitions}
This section gives basic definitions, which are based on exactly-1 disjunction, denoted by $\X$.

\begin{definition}
A literal $\lt_{i\hskip-0.12em}$ is a variable $x_{i\hskip-0.09em}$ assigned true, or a negated variable $\nx{i\hskip-0.09em}$ assigned true. That is, $\lt_{i\hskip-0.10em} \in \{x_i, \nx{i}\}$, in which $x_{i\hskip-0.12em} = \mathbf{T}$ and $\nx{i\hskip-0.12em} = \mathbf{T}$.
\end{definition}

\begin{definition}
A clause $C_{k\hskip-0.12em} = (\lt_{i\hskip-0.07em} \X \lt_{j\hskip-0.05em} \X \lt_u)$ denotes an exactly-1 disjunction of literals.
\end{definition}

\begin{definition}\label{d:minterm}
$c_{k\hskip-0.09em} = \hskip-0.15em\bigwedge\hskip-0.15em \lt_{i\hskip-0.12em}$ denotes a \emph{minterm}, a conjunction of $\lt_{i\hskip-0.05em}$, where $\lt_{i\hskip-0.12em}$ is called a \emph{conjunct}.
\end{definition}

\begin{definition}\label{d:formula}
$\FM =  \scp \wedge \Fm$ denotes an \emph{X3SAT} formula such that $\scp = \hskip-0.11em\bigwedge\hskip-0.11em c_{k\hskip-0.09em}$ and $\Fm = \hskip-0.11em\bigwedge\hskip-0.11em C_{k\hskip-0.05em}$.
\end{definition}

Where appropriate, $C_{k\hskip-0.05em}$, as well as $\scp$, is denoted by a set. Thus, $\FM = \scp \wedge \Fm$ the formula, that is, $\FM = \scp \wedge C_{1\hskip-0.09em} \wedge C_{2\hskip-0.05em} \wedge \cdots \wedge C_{m\hskip-0.05em}$, is denoted by $\FM = \{\scp, C_1, C_2, \ldots, C_m\}$ the family of sets.

\begin{definition}\label{d:SatCk}
$C_{k\hskip-0.09em} = (\lt_{i\hskip-0.09em} \X \lt_{j\hskip-0.09em} \X \lt_u)$ is satisfied iff $(\lt_{i\hskip-0.07em} \wedge \nlt_{j\hskip-0.07em} \wedge \nlt_u) \vee (\nlt_{i\hskip-0.07em} \wedge \lt_{j\hskip-0.07em} \wedge \nlt_u) \vee (\nlt_{i\hskip-0.07em} \wedge \nlt_{j\hskip-0.07em} \wedge \lt_u)$ is satisfied, since any clause $C_{k\hskip-0.09em}$ contains \emph{exactly one} true literal by the definition of \emph{X3SAT}.
\end{definition}

\begin{definition}[Incompatibility]\label{d:Incm}
$\lt_{i\hskip-0.12em}$ in some $C_{k\hskip-0.09em}$ is incompatible, denoted by $\neg \lt_{i\hskip-0.05em}$, iff $\lt_{i\hskip-0.09em}$ leads to a contradiction $x_{j\hskip-0.09em} \wedge \nx{j\hskip-0.05em}$, that is, $\lt_{i\hskip-0.07em} \wedge \FM$ is unsatisfiable, hence $\lt_{i\hskip-0.12em}$ is removed from every $C_{k\hskip-0.09em}$ in $\Fm$.
\end{definition}

\begin{remark*}
Each $x_{i\hskip-0.09em}$ and $\nx{i\hskip-0.09em}$ in $\Fm$ is assumed to be compatible, thus no $C_{k\hskip-0.09em}$ contains $\neg x_{i\hskip-0.05em}$, or $\neg \nx{i\hskip-0.05em}$, while any $\lt_{i\hskip-0.12em}$ in $\scp$ is \emph{necessarily} true by Definition \ref{d:minterm}/\ref{d:formula}, thus denotes a \emph{conjunct}, to satisfy $\FM$.
\end{remark*}

\begin{note}\label{n:incmp}
If $\lt_{i\hskip-0.12em} \in \scp$, then $\lt_{i\hskip-0.09em} \imp \neg \nlt_{i\hskip-0.05em}$, that is, $\nlt_{i\hskip-0.12em}$ becomes incompatible, and is \emph{removed} from $\Fm$. If $\lt_{i\hskip-0.09em} \imp x_{j\hskip-0.09em} \wedge \nx{j\hskip-0.05em}$, hence $\neg x_{j\hskip-0.12em} \vee \neg \nx{j\hskip-0.12em} \imp \neg \lt_{i\hskip-0.05em}$, then $\neg \lt_{i\hskip-0.09em} \imp \nlt_{i\hskip-0.05em}$, that is, $\nlt_{i\hskip-0.12em}$ becomes a \emph{conjunct} ($\nlt_{i\hskip-0.12em} \in \scp$).
\end{note}

\begin{definition}
$\Li = \{1, 2, \ldots, n\}$ denotes the index set of the literals $\lt_{i\hskip-0.05em}$, $\Ckk = \{1, 2, \ldots, m\}$ denotes the index set of the clauses $C_{k\hskip-0.05em}$, and $\Ckk^{\lt_i} \!= \{k \in \Ckk \,|\, \lt_{i\hskip-0.09em} \in C_k\}$ denotes $C_{k\hskip-0.09em}$ containing $\lt_{i\hskip-0.05em}$.
\end{definition}

\begin{example}
$\FM \hskip-0.03em= \nx{4\hskip-0.04em} \wedge (x_{1\hskip-0.12em} \X \nx{2\hskip-0.04em} \X x_3) \wedge (\nx{3\hskip-0.04em} \X \nx4)$, in which $\nx{4\hskip-0.09em}$ is necessary for satisfying $\FM$, thus $\scp \hskip-0.03em= \{\nx4\}$, $\Ckk^{\nx{4\hskip-0.11em}} = \{2\}$, and $C_{1\hskip-0.12em} = \{x_{1\hskip-0.07em}, \nx2, x_3\}$ denotes \emph{either} $x_{1\hskip-0.12em} = \mathbf{T}$ \emph{or} $\nx{2\hskip-0.05em} = \mathbf{T}$ \emph{or} $x_{3\hskip-0.05em} = \mathbf{T}$.
\end{example}

\begin{definition}[Collapse]\label{d:colp}
A clause $C_{k\hskip-0.09em} = (\lt_{i\hskip-0.09em} \X x_{j\hskip-0.09em} \X \nx u)$ is said to \emph{collapse} to the minterm $c_{k\hskip-0.09em} = (\lt_{i\hskip-0.07em} \wedge \nx{j\hskip-0.09em} \wedge x_u)$, thus $\lt_{i\hskip-0.09em} \notin C_{k\hskip-0.05em}$, if $\lt_{i\hskip-0.12em}$ is \emph{necessary}, denoted by $(\lt_{i\hskip-0.09em} \X x_{j\hskip-0.09em} \X \nx u) \clp (\lt_{i\hskip-0.07em} \wedge \nx{j\hskip-0.09em} \wedge x_u)$.
\end{definition}

\begin{definition}[Shrinkage]\label{d:shrn}
A clause $C_{k\hskip-0.09em} = (\lt_{i\hskip-0.09em} \X \lt_{j\hskip-0.09em} \X \lt_u)$ is said to shrink to another clause $C_{k'\hskip-0.09em} = (\lt_{j\hskip-0.09em} \X \lt_u)$, if $\neg \lt_{i\hskip-0.12em}$ $(\lt_{i\hskip-0.12em}$ \emph{the incompatible is removed)}, denoted by $(\lt_{i\hskip-0.09em} \X \lt_{j\hskip-0.09em} \X \lt_u) \rdc (\lt_{j\hskip-0.09em} \X \lt_u)$.
\end{definition}

\begin{definition}[Compatibility of $\lt_{i\hskip-0.10em} \in \{x_i, \nx{i}\}$ over $\Fm$] \label{d:Fmri}
$\Fm(\lt_i) = \lt_{i\hskip-0.07em} \wedge \Fm$ for any $\lt_{i\hskip-0.09em} \in C_{k\hskip-0.09em}$ in $\Fm$.
\end{definition}

\begin{note}[Reduction]\label{n:rdctn}
The collapse or shrinkage denotes a reduction. If $\lt_{i\hskip-0.12em} \in \scp$, then $\lt_{i\hskip-0.12em}$ leads to \emph{reductions} over $\Fm$, thus $\FM \Rdc \FM'\hskip-0.09em$. That is, $\FM \Rdc \FM'\hskip-0.09em$ iff $C_{k\hskip-0.09em} \clp \,c_{k\hskip-0.09em}$ or $C_{k\hskip-0.09em} \rdc C_{k'\hskip-0.12em}$ for $C_{k\hskip-0.12em}$ in $\Fm$. Since $\lt_{i\hskip-0.12em}$ is necessary for $\Fm(\lt_i)$, it leads to \emph{reductions} over $\Fm(\lt_i)$. Then, $(\nlt_{i\hskip-0.07em} \X \lt_{v\hskip-0.09em} \X \lt_y) \rdc (\lt_{v\hskip-0.09em} \X \lt_y)$ and $(\lt_{i\hskip-0.07em} \X x_{j\hskip-0.09em} \X \nx u) \clp (\lt_{i\hskip-0.05em} \wedge \nx{j\hskip-0.09em} \wedge x_u)$, because $\lt_{i\hskip-0.07em} \imp \hskip-0.05em \neg \nlt_{i\hskip-0.09em}$ such that $\lt_{i\hskip-0.07em} \imp \lt_{i\hskip-0.07em} \wedge \nx{j\hskip-0.09em} \wedge x_{u\hskip-0.09em}$ holds over some $C_{k\hskip-0.09em} = (\lt_{i\hskip-0.07em} \X x_{j\hskip-0.09em} \X \nx u)$, since $\lt_{i\hskip-0.05em} \imp \neg x_{j\hskip-0.09em} \wedge \neg \nx{u\hskip-0.05em}$, thus $\neg x_{j\hskip-0.09em} \imp \nx{j\hskip-0.09em}$ and $\neg \nx{u\hskip-0.09em} \imp x_{u\hskip-0.09em}$ (see Definition \ref{d:SatCk}/\ref{d:Incm}).
\end{note}

\begin{definition}\label{d:GnFm}
$\Fm$ denotes a general formula if $\{x_i, \nx{i}\} \nsubseteq C_{k\hskip-0.09em}$ for any $i \in \Li$ and $k \in \Ckk$, hence $\Ckk^{x_i\hskip-0.09em} \cap \Ckk^{\nx{i}\hskip-0.09em} = \emptyset$. $\Fm$ denotes a \emph{special} formula if $\{x_i, \nx{i}\} \subseteq C_{k\hskip-0.09em}$ for some $k$, hence $\Ckk^{x_i\hskip-0.09em} \cap \Ckk^{\nx{i}\hskip-0.09em} = \{k\}$.
\end{definition}

\begin{lemma}[Conversion of a special formula]\label{l:Cnvrt}
Each \emph{clause} $C_{k\hskip-0.09em} = (\lt_{j\hskip-0.09em} \X  x_{i\hskip-0.09em} \X \nx i)$ is replaced by the \emph{conjunct} $\nlt_{j\hskip-0.12em}$ so that $\Ckk^{x_i\hskip-0.09em} \cap \Ckk^{\nx{i}\hskip-0.09em} = \emptyset$ for any $i \in \Li$, if $\Fm = \hskip-0.11em\bigwedge\hskip-0.11em C_{k\hskip-0.12em}$ is a special formula.
\end{lemma}
\begin{proof}
$\Fm$ is unsatisfiable due to $\lt_{j\hskip-0.09em} \imp \nx{i\hskip-0.07em} \wedge x_{i\hskip-0.05em}$. Then, $x_{i\hskip-0.12em} \vee \hskip0.05em\nx{i\hskip-0.09em} \imp \nlt_{j\hskip-0.05em}$. That is, $\nlt_{j\hskip-0.12em}$ is \emph{necessary} for satisfying $C_{k\hskip-0.09em} = (\lt_{j\hskip-0.09em} \X  x_{i\hskip-0.09em} \X \nx i)$, which is sufficient also, thus $\nlt_{j\hskip-0.12em}$ is equivalent to $C_{k\hskip-0.05em}$. Therefore, each \emph{clause} $C_{k\hskip-0.09em} = (\lt_{j\hskip-0.09em} \X  x_{i\hskip-0.09em} \X \nx i)$ is replaced by the \emph{conjunct} $\nlt_{j\hskip-0.09em}$ so that $\Ckk^{x_i\hskip-0.09em} \cap \Ckk^{\nx{i}\hskip-0.09em} = \emptyset$.
\end{proof}

\begin{example}
$\FM = (x_{2\hskip-0.05em} \X \nx1) \wedge (x_{1\hskip-0.14em} \X \nx{3\hskip-0.05em} \X x_4) \wedge (x_{1\hskip-0.14em} \X \nx{2\hskip-0.05em} \X x_2)$ is a special formula due to $C_{3\hskip-0.09em} = \{x_{1\hskip-0.07em}, \nx2, x_2\}$. Note that $\Ckk^{\nx{2}\hskip-0.09em} \cap \Ckk^{x_2\hskip-0.09em} = \{3\}$. Then, $\FM$ is converted by replacing the clause $C_{3\hskip-0.12em}$ with the conjunct $\nx{1\hskip-0.07em}$. As a result, $\FM \gets \nx{1\hskip-0.12em} \wedge (x_{2\hskip-0.05em} \X \nx1) \wedge (x_{1\hskip-0.12em} \X \nx3 \X x_4)$. Likewise, if $\FM = \linebreak (x_{3\hskip-0.05em} \X \nx{4\hskip-0.05em} \X x_4) \wedge (\nx{3\hskip-0.05em} \X x_{2\hskip-0.05em} \X \nx2) \wedge (x_{2\hskip-0.05em} \X \nx1)$, then $\FM \gets \nx{3\hskip-0.05em} \wedge x_{3\hskip-0.05em} \wedge (x_{2\hskip-0.05em} \X \nx1)$, which is unsatisfiable.
\end{example}
\section{\texorpdfstring{The $\FM$ Scan}{The Formula Scan}}
The $\FM$ scan asserts that $\FM$ is satisfiable iff $x_{i\hskip-0.09em}$ or $\nx{i\hskip-0.12em}$ is compatible (Definition \ref{d:Fmri}) for all $i \in \Li$. Hence, we need to show that $\Fm(x_1)$ or $\Fm(\nx1)$, and $\Fm(x_2)$ or $\Fm(\nx2)$, and $\cdots$ and $\Fm(x_n)$ or $\Fm(\nx n)$ are satisfied. If $\FM$ is satisfiable, then a satisfying assignment is determined (see Section \ref{s:DetAsg}).

$\unsat \FM$ denotes $\FM$ is unsatisfiable, and $\sat \FM$ denotes that $\alpha = \{\lt_{1\hskip-0.05em}, \lt_2, \ldots, \lt_n\}$ is a satisfying assignment for $\FM$. $\scp \ent \scp'\hskip-0.07em$ denotes that $\scp$ \emph{entails} $\scp'\hskip-0.07em$, and $\scp \prv \scp'\hskip-0.07em$ denotes that $\scp$ \emph{proves} $\scp'\hskip-0.07em$.

$\FM_{s\hskip-0.07em}$ for $s \GE 2$ denotes the \emph{current} formula at the $s$\textsuperscript{th} scan/step such that $\FM \coloneqq \FM_1\hskip-0.1em$, after $\neg \lt_{j\hskip-0.09em}$ holds in $\Fm_{s - 1\hskip-0.1em}$ (see Definition \ref{d:Incm}). Then, $\Fm_s^{\lt_i} \!= (\lt_{ik_1\!\hskip-0.09em} \X \lt_{u_1k_1\!\hskip-0.07em} \X \lt_{u_2k_1}) \wedge \cdots \wedge (\lt_{ik_r\!\hskip-0.03em} \X \lt_{v_1k_r\!\hskip-0.03em} \X \lt_{v_2k_r})$ denotes the formula over clauses $C_{k\hskip-0.07em} \ni \lt_{i\hskip-0.09em}$ in $\Fm_{s\hskip-0.05em}$, where $\lt_{i\hskip-0.09em} \in \{x_i, \nx{i}\}$. Hence, $\Ckk_s^{\lt_i} \!= \{k_{1\hskip-0.07em}, \ldots, k_r\}$.

$\lcl{s}(\lt_i)$ is called the \emph{local} effect of $\lt_{i\hskip-0.05em}$, and $\lcln_s(\neg \lt_i)$ is the effect of $\neg \lt_{i\hskip-0.05em}$. $\ovrl_s(\lt_i)$ denotes its \emph{overall} effect such that $\ovrl_s(\lt_i) = \lcl{s}(\lt_i) \wedge \lcln_s(\neg \nlt_i)$, specified below. Also, $\lcl{s}(\lt_i) = \hskip-0.11em\bigwedge\hskip-0.03em (c_{k\hskip-0.07em} \wedge C_{k\hskip-0.07em})$ such that $|C_k| = 1$. Moreover, $\lcln_s(\neg \lt_i) = \hskip-0.11em\bigwedge\hskip-0.11em C_{k\hskip-0.09em}$ such that $|C_k| > 1$, or $\lcln_s(\neg \lt_i)$ is empty.

\subsection[Incompatibility-Reductions]{Introduction: Incompatibility and Reductions}
Example \ref{e:Ex1} (\ref{e:Ex2}) introduces incompatibility (reductions over $\Fm$), which drive the $\FM$ scan.

\begin{example}\label{e:Ex1}
Consider $\Fm(x_1)$ over ${\color{purple}\FM = \Fm} = (x_{1\hskip-0.12em} \X \nx3) \wedge (x_{1\hskip-0.12em} \X \nx{2\hskip-0.05em} \X x_3) \wedge (x_{2\hskip-0.05em} \X \nx3)$. Thus, $x_{1\hskip-0.12em}$ is necessary for $\Fm(x_1)$, hence $x_{1\hskip-0.09em} \ent \lcl{}(x_1)$ such that $\lcl{}(x_1) = (x_{1\hskip-0.10em} \wedge x_3) \wedge (x_{1\hskip-0.10em} \wedge {\color{white!30!blue}x_{2\hskip-0.04em}} \wedge \nx3)$. That is, $x_{1\hskip-0.07em} \imp \neg \nx{3\hskip-0.07em}$ holds over $C_{1\!} = (x_{1\hskip-0.12em} \X \nx3)$, hence $\neg \nx{3\hskip-0.09em} \imp x_{3\hskip-0.05em}$. Likewise, $x_{1\hskip-0.07em} \imp \neg \nx{2\hskip-0.05em} \wedge \neg x_{3\hskip-0.07em}$ holds over $(x_{1\hskip-0.12em} \X \nx{2\hskip-0.05em} \X x_3)$, hence $\neg \nx{2\hskip-0.09em} \imp \color{white!30!blue}x_{2\hskip-0.05em}$ and $\neg x_{3\hskip-0.09em} \imp \nx{3\hskip-0.05em}$ (see Note \ref{n:rdctn}). Thus, $\ovrl(x_1) = \lcl{}(x_1) \wedge \lcln(\neg \nx1)$ becomes the overall effect, where $\lcln(\neg \nx1)$ is empty. Then, the reductions initiated by $x_{1\hskip-0.12em}$ over $\Fm(x_1)$ are to proceed due to $\color{white!30!blue}x_2$. Nevertheless, they are interrupted by $x_{3\hskip-0.05em} \wedge \nx3$ due to $\lcl{}(x_1)$. Hence, $\Fm(x_1) = \ovrl(x_1) \wedge (x_{2\hskip-0.05em} \X \nx3)$ is unsatisfiable, thus $x_{1\hskip-0.12em}$ is \emph{incompatible} for $\FM$, i.e, $\neg x_{1\hskip-0.07em} \imp \nx{1\hskip-0.07em}$.
\end{example}

\begin{example}\label{e:Ex2}
$\nx{1\hskip-0.12em}$ initiates \emph{reductions} over $\Fm$ (Note \ref{n:rdctn}). Then, $\lcl{}(\nx1) \hskip-0.05em= \nx{1\hskip-0.12em}\wedge \nx3$, $\lcln(\neg x_1) \hskip-0.05em= (\nx{2\hskip-0.05em} \X x_3)$, and $\ovrl(\nx1) \hskip-0.05em= \lcl{}(\nx1) \wedge \lcln(\neg x_1)$ to define ${\color{purple}\FM_2} \hskip-0.11em= \ovrl(\nx1) \wedge (x_{2\hskip-0.05em} \X \nx3)$. Note that $(x_{2\hskip-0.05em} \X \nx3)$ is beyond $\ovrl(\nx1)$ the overall effect. Note also that $\{\nx3\} \notin \lcln(\neg x_1)$, while $\nx{3\hskip-0.07em} \in \lcl{}(\nx1)$, because $C_{1\!} \rdc c_{1\hskip-0.07em}$, since $\lcln(\neg x_1)$ contains no singleton. Then, $\FM_2$ is the current formula due to the first reduction by  $\nx{1\hskip-0.12em}$ over $\Fm$. Thus, $\FM \Rdc \FM_2$ due to $(x_{1\hskip-0.12em} \X \nx3) \rdc (\nx3)$ and $(x_{1\hskip-0.12em} \X \nx{2\hskip-0.07em} \X x_3) \rdc (\nx{2\hskip-0.07em} \X x_3)$. As a result, ${\color{purple}\FM_2} \hskip-0.11em= \nx{1\hskip-0.09em} \wedge \nx{3\hskip-0.03em} \wedge (\nx{2\hskip-0.07em} \X x_3) \wedge (x_{2\hskip-0.09em} \X \nx3)$, in which $\scp_{2\hskip-0.07em} = \{\nx{1\hskip-0.07em}, \nx3\}$ denotes the conjuncts, and $C_{1\hskip-0.07em} = \{\nx2, x_3\}$ and $C_2 = \{x_2, \nx3\}$ denote the clauses. Note that $\Ckk_2^{x_3\hskip-0.09em} = \{1\}$ and $\Ckk_2^{\nx3\hskip-0.09em} = \{2\}$. Likewise, $\nx{3\hskip-0.07em}$ leads to the next reduction over $\Fm_2$: $\lcl{2}(\nx3) = (\nx{2\hskip-0.05em} \wedge \nx3)$, $\lcln_2(\neg x_3)$ is empty, and $\ovrl_2(\nx3) \hskip-0.07em= \lcl{2}(\nx3) \hskip-0.07em\wedge\hskip-0.05em \lcln_2(\neg x_3)$. Thus, $\FM_2 \hskip-0.05em\Rdc \FM_3$ due to $(x_{2\hskip-0.09em} \X \nx3) \hskip-0.07em\clp (\nx{2\hskip-0.07em} \wedge \nx3)$ and $(\nx{2\hskip-0.09em} \X x_3) \hskip-0.07em\rdc (\nx2)$. Then, ${\color{purple}\FM_3} = \ovrl(\nx1) \wedge \ovrl_2(\nx3) = \nx{1\hskip-0.12em} \wedge \nx2 \wedge \nx3$, which denotes the cumulative effects of $\nx{1\hskip-0.12em}$ and $\nx{3\hskip-0.05em}$.
\end{example}

\subsection{\texorpdfstring{The Core Algorithms: \texttt{Scope} and \texttt{Scan}}{The core algorithms}}\label{s:algrthms}
Let $\Fm_s^{\lt_j} \!= (\lt_{jk_1\hskip-0.2em} \X \lt_{i_1k_1\hskip-0.2em} \X \lt_{i_2k_1}) \wedge \cdots \wedge (\lt_{jk_r\!} \X \lt_{u_1k_r\!} \X \lt_{u_2k_r})$ for Lemma \ref{l:lclocr} and \ref{l:lclNocr} below.

\begin{lemma}\label{l:lclocr}
$\lt_{j\hskip-0.09em} \ent \lcl{s}(\lt_j)$ such that $\lcl{s}(\lt_j) = \lt_{j\hskip-0.09em} \wedge \nlt_{i_1\hskip-0.15em} \wedge \nlt_{i_2\hskip-0.09em} \wedge \cdots \wedge \nlt_{u_1\hskip-0.15em} \wedge \nlt_{u_2\hskip-0.07em}$, unless $\unsat \lcl{s}(\lt_j)$.
\end{lemma}
\begin{proof}
Follows from Definition \ref{d:colp}. That is, $\lt_{j\hskip-0.09em} \imp (\lt_{j\hskip-0.09em} \wedge \nlt_{i_1\hskip-0.15em} \wedge \nlt_{i_2}) \wedge \cdots \wedge (\lt_{j\hskip-0.09em} \wedge \nlt_{u_1\hskip-0.15em} \wedge \nlt_{u_2})$. Hence, $\lt_{j\hskip-0.09em} \imp \lt_{j\hskip-0.09em} \wedge \nlt_{i_1\hskip-0.15em} \wedge \nlt_{i_2\hskip-0.09em} \wedge \cdots \wedge \nlt_{u_1\hskip-0.15em} \wedge \nlt_{u_2\hskip-0.07em}$.
\end{proof}

\begin{lemma}\label{l:lclNocr}
If $\neg \lt_{j\hskip-0.05em}$, then $\lcln_s(\neg \lt_j)$ holds such that $\lcln_s(\neg \lt_j) = (\lt_{i_1\hskip-0.19em} \X \lt_{i_2}) \wedge \cdots \wedge (\lt_{u_1\hskip-0.19em} \X \lt_{u_2})$.
\end{lemma}
\begin{proof}
Follows from Definition \ref{d:shrn}. $\lcln_s(\neg \lt_j) = \big\{\{\}\big\}$, or $|C_k| > 1$ for any $C_{k\hskip-0.09em}$ in $\lcln_s(\neg \lt_j)$.
\end{proof}

\begin{lemma}[Overall effect of $\lt_{j\hskip-0.09em}$ over $\Fm_s$]\label{l:ovrllefc}
$\ovrl_s(\lt_j) = \lcl{s}(\lt_j) \wedge \lcln_s(\neg \nlt_j)$.
\end{lemma}
\begin{proof}
Follows from Lemma \ref{l:lclocr}, and from \ref{l:lclNocr} via $\Fm_s^{\nlt_j}\!$, since $\lt_{j\hskip-0.09em} \imp \neg \nlt_{j\hskip-0.05em}$, thus $\lt_{j\hskip-0.09em} \ent \lt_{j\hskip-0.09em} \wedge \neg \nlt_{j\hskip-0.05em}$.
\end{proof}

The algorithm \texttt{OvrlEft}~$\!(\lt_j, \Fm_*)$ below constructs the overall effect $\ovrl_*(\lt_j)$ by means of the local effect $\lcl{*}(\lt_j)$ (see Lines 1-6, or L:1-6), as well as of the local effect $\lcln_*(\neg \nlt_j)$ (L:7-10).

\begin{algorithm}\caption{\texttt{OvrlEft}~$\!(\lt_j, \Fm_*)$ \Cmmnt{Construction of the overall effect $\ovrl_*(\lt_j)$ due to Lemma \ref{l:ovrllefc}}}
\begin{algorithmic}[1]
\ForAll {$k \in \Ckk_*^{\lt_j\hskip-0.12em}$ over $\Fm_*$} \Cmmnt{Construction of the local effect $\lcl{*}(\lt_j)$ due to $\lt_{j\hskip-0.12em}$ (Lemma \ref{l:lclocr})}
    \ForAll {$\lt_{i\hskip-0.09em} \in \big(C_k - \{\lt_j\}\big)$}\Cmmnt{$\lcl{*}(\lt_j)$ gets $\lt_{j\hskip-0.12em}$ via $\lt_e$ (see \texttt{Scope} \hyperref[alg:Scope]{L:4}), or via $\nlt_{j\hskip-0.12em}$ (\texttt{Remove} \hyperref[alg:Remove]{L:2})}
        \State $c_{k\hskip-0.09em} \gets c_{k\hskip-0.09em} \cup \{\nlt_i\}$; \Cmmnt{$(\lt_{jk\hskip-0.09em} \X \lt_{i_1k\hskip-0.07em} \X \lt_{i_2k}) \clp (\nlt_{i_1k\hskip-0.07em} \wedge \nlt_{i_2k})$. That is, $C_{k\hskip-0.09em} \clp \,c_{k\hskip-0.09em}$ (see Definition \ref{d:minterm}/\ref{d:colp})}
    \EndFor
    \State $\color{green!40!black}\lcl{*}(\lt_j) \gets \lcl{*}(\lt_j) \cup c_{k\hskip-0.05em};$ \Cmmnt{$c_{k\hskip-0.09em}$ consists in $\scp_s(\lt_j)$ (see \texttt{Scope} \hyperref[alg:Scope]{L:4}), or in $\scp_{s\hskip-0.07em}$ (see \texttt{Remove} \hyperref[alg:Remove]{L:2})}
\EndFor \Cmmnt{L:1-6 are independent from L:7-10, since $\Ckk_*^{\lt_j} \!\cap \Ckk_*^{\nlt_j} \!= \emptyset$, i.e., $\Ckk_*^{x_j} \!\cap \Ckk_*^{\nx j} \!= \emptyset$ (Lemma \ref{l:Cnvrt})}
\ForAll {$k \in \Ckk_*^{\nlt_j\hskip-0.12em}$ over $\Fm_*$}  \Cmmnt{Construction of the local effect $\lcln_*(\neg \nlt_j)$ due to $\neg \nlt_{j\hskip-0.12em}$ (Lemma \ref{l:lclNocr})}
    \State $C_k \gets C_k - \{\nlt_j\}$; \!\Cmmnt{$(\nlt_{jk\hskip-0.09em} \X \lt_{u_1k\hskip-0.09em} \X \lt_{u_2k}) \rdc (\lt_{u_1k\hskip-0.09em} \X \lt_{u_2k})$ or $(\nlt_{jk\hskip-0.09em} \X \lt_{uk}) \rdc (\lt_{uk})$ (Definition \ref{d:shrn})}
    \State \textbf{if} $|C_k| = 1$ \textbf{then} $\color{green!40!black}\lcl{*}(\lt_j) \gets \lcl{*}(\lt_j) \cup C_{k\hskip-0.05em};$ $C_{k\hskip-0.09em} \gets \emptyset$; \!\Cmmnt{$\!\lcln_*(\neg \nlt_j)$ contains no singleton, $C_{k\hskip-0.09em} \rdc c_{k\hskip-0.09em}$}
\EndFor \Cmmnt{3$\backslash$2-literal $C_{k\hskip-0.09em}$ in $\Fm_*^{\nlt_{j\hskip-0.12em}}$ shrinks due to $\neg \nlt_{j\hskip-0.12em}$ to 2-literal $C_{k\hskip-0.09em}$ in $\Fm_*^{\nlt_{j\hskip-0.09em}}\backslash$to \emph{conjunct} $\lt_{u\hskip-0.09em}$ in $\lcl{*}(\lt_j)$}
\color{white!30!blue} \State \Return $\lcl{*}(\lt_j)$ \& $\lcln_*(\neg \nlt_j) \gets \Fm_*^{\nlt_j}$; \!\Cmmnt{$\lcl{*}(\lt_j) \hskip-0.03em= \hskip-0.11em\bigwedge\hskip-0.03em (c_{k\hskip-0.05em} \wedge C_k)$, $|C_k| = 1$ \& $\lcln_*(\neg \nlt_j) \hskip-0.03em= \hskip-0.11em\bigwedge\hskip-0.11em C_{k\hskip-0.05em}$, $|C_k| > 1$}
\end{algorithmic}
\label{alg:Ovrl}
\end{algorithm}

$\scp_s(\lt_j)$ is called the scope of $\lt_{j\hskip-0.05em}$, and $\Fm'_s(\lt_j)$ is called beyond the scope, defined over $\Fm_{s\hskip-0.05em}$.

\begin{lemma}[Scope of $\lt_j$]\label{l:Scope}
$\lt_{j\hskip-0.12em}$ transforms $\Fm_{s\hskip-0.12em}$ into $\Fm_s(\lt_j) \hskip-0.07em= \scp_s(\lt_j) \wedge \Fm'_s(\lt_j)$, unless $\unsat \scp_s(\lt_j)$, where $\scp_s(\lt_j) =  \lt_{j\hskip-0.09em} \wedge \lt_{i\hskip-0.09em} \wedge \cdots \wedge \lt_{u\hskip-0.09em}$ and $\Fm'_s(\lt_j) = \hskip-0.11em\bigwedge\hskip-0.11em C_{k\hskip-0.05em}$. Thus, $\lt_{j\hskip-0.09em} \ent \scp_s(\lt_j)$, hence $\lt_{j\hskip-0.09em} \prv \scp_s(\lt_j)$.
\end{lemma}
\begin{proof}
$\Fm_s(\lt_j) \hskip-0.09em= \lt_{j\hskip-0.09em} \wedge \Fm_{s\hskip-0.11em}$ by Definition \ref{d:Fmri}. Then, $\lt_{j\hskip-0.12em}$ initiates a \emph{deterministic} chain of reductions (see Note \ref{n:rdctn}). As a result, $\lt_{j\hskip-0.09em} \imp \lt_{j\hskip-0.09em} \wedge x_{i\hskip-0.05em} \wedge \nx{u\hskip-0.12em}$ holds over each $C_{k\hskip-0.09em} = (\lt_{j\hskip-0.12em} \X \nx{i\hskip-0.05em} \X x_u)$ containing $\lt_{j\hskip-0.05em}$, and $\neg \nlt_{j\hskip-0.12em} \imp (\nx{u\hskip-0.09em} \X x_{v\hskip-0.05em})$ holds over each $C_{k\hskip-0.09em} = (\nlt_{j\hskip-0.09em} \X \nx{u\hskip-0.05em} \X x_{v\hskip-0.03em})$ containing $\nlt_{j\hskip-0.05em}$. These reductions thus proceed, as long as new conjuncts $\lt_{e\hskip-0.09em}$ emerge in $\Fm_s(\lt_j)$ (see \texttt{Scope} L:2-4). If the reductions are interrupted, then $\lt_{j\hskip-0.12em}$ is incompatible (L:5). If they terminate, then $\scp_s(\lt_j)$ and $\Fm'_s(\lt_j)$ are constructed (L:9). Thus, $\lt_{j\hskip-0.09em} \ent \scp_s(\lt_j)$. It is obvious that if $\lt_{j\hskip-0.09em} \ent \scp_s(\lt_j)$, then $\lt_{j\hskip-0.09em} \prv \scp_s(\lt_j)$.
\end{proof}

\begin{algorithm}\caption{\texttt{Scope}~$\!(\lt_j, \Fm_s)$ \Cmmnt{Construction of $\scp_s(\lt_j)$ and $\Fm'_s(\lt_j)$ due to $\lt_{j\hskip-0.09em}$ over $\Fm_s$; $\FM_{s\hskip-0.07em} = \scp_{s\hskip-0.07em} \wedge \Fm_{s\hskip-0.07em}$}}
\begin{algorithmic}[1]
\State $\scp_s(\lt_j) \gets \{\lt_j\}$;  $\Fm_* \gets \Fm_s$; \Cmmnt{$\Fm_s(\lt_j) \coloneqq \lt_{j\hskip-0.09em} \wedge \Fm_{s\hskip-0.05em}$. $\scp_{s\hskip-0.07em}$ and $\Fm_{s\hskip-0.07em}$ are disjoint due to \texttt{Scan} \hyperref[alg:Scan]{L:1-3}}
    \ForAll {$\lt_{e\hskip-0.09em} \in \big(\scp_s(\lt_j) - R\big)$} \Cmmnt{Reductions of $C_{k\hskip-0.09em}$ initiated by $\lt_{j\hskip-0.12em}$ over $\Fm_s$ start off}
        \State \texttt{OvrlEft}~$\!(\lt_e, \Fm_*)$; \Cmmnt{It returns $\lcl{*}(\lt_e)$ for L:4 \& $\lcln_*(\neg \nlt_e)$ for L:6}
        \State $\scp_s(\lt_j) \hskip-0.09em \gets \scp_s(\lt_j) \cup \{\lt_e\} \cup \color{green!40!black}\lcl{*}(\lt_e);$ \!\Cmmnt{\!$\color{green!40!black}\lcl{*}(\lt_e)$ due to \texttt{OvrlEft} L:5,9 consists in the scope $\scp_s(\lt_j)$\!\!} \color{red}
        \State \textbf{if} $\scp_s(\lt_j) \supseteq \{x_i, \nx i\}$ \textbf{then} \Return NULL; \Cmmnt{$\lt_{j\hskip-0.09em} \imp x_{i\hskip-0.07em} \wedge \nx{i\hskip-0.05em}$, $i \in \Li^{_\Fm}\!$. $\unsat \scp_s(\lt_j)$, thus $\unsat \Fm_s(\lt_j)$} \color{black}
        \State $\lcln_*(\neg\lt) \gets \lcln_*(\neg\lt) \cup \lcln_*(\neg \nlt_e)$; \!\Cmmnt{\!$\lcln_*(\neg\lt) \hskip-0.09em=\hskip-0.05em \big\{\{\}\big\}$ or $\lcln_*(\neg\lt) \hskip-0.09em=\hskip-0.05em \bigcup C_{k\hskip-0.05em}, \hskip0.02em|C_k| \hskip-0.09em>\hskip-0.09em 1$ (\texttt{OvrlEft} L:8-11)}
        \State $\Fm_* \gets \lcln_*(\neg\lt) \wedge \Fm'_*$; $R \gets R \cup \{\lt_e\}$; \Cmmnt{$\lcln_*(\neg\lt)$ and $\Fm'_*\hskip-0.09em$ consist in beyond the scope $\Fm'_s(\lt_j)$} \linebreak
        \Cmmnt{$\Fm'_*\hskip-0.09em = \hskip-0.11em\bigwedge\hskip-0.11em C_{k\hskip-0.09em}$ for $k \in \Ckk'_*$, where $\Ckk'_* \!= \Ckk_* - (\Ckk_*^{x_e\hskip-0.09em} \cup \Ckk_*^{\nx e})$, and $\Ckk_*^{x_e\hskip-0.09em} \cap \Ckk_*^{\nx e} = \emptyset$ due to Lemma \ref{l:Cnvrt}}
    \EndFor \Cmmnt{The reductions terminate if $\scp_s(\lt_j) = R$, which denotes conjuncts already reduced $C_{k\hskip-0.12em}$}
\color{white!30!blue} \State \Return $\scp_s(\lt_j)$ \& $\Fm'_s(\lt_j) \gets \Fm_*$; \Cmmnt{$\Fm_s(\lt_j) \hskip-0.03em= \scp_s(\lt_j) \wedge \Fm'_s(\lt_j)$. $\scp_s(\lt_j) \hskip-0.03em= \hskip-0.15em\bigwedge\hskip-0.15em \lt_{j\hskip-0.12em}$ and $\Fm'_s(\lt_j) \hskip-0.03em= \hskip-0.11em\bigwedge\hskip-0.11em C_{k\hskip-0.07em}$}
\end{algorithmic}
\label{alg:Scope}
\end{algorithm}

\begin{note}\label{n:ScpTrmnts}
$\Li_s(\lt_j)$ being an index set of $\scp_s(\lt_j)$, $\Li_s(\lt_j) \cap \Li'_s(\lt_j) = \emptyset$ and $\Li_s(\lt_j) \cup \Li'_s(\lt_j) = \Li^{_\Fm\!}$, if \texttt{Scope}~$\!(\lt_j, \Fm_s)$ terminates. Thus, $\scp_s(\lt_j)$ and $\Fm'_s(\lt_j)$ are disjoint, where $\Fm'_s(\lt_j)$ can be empty.
\end{note}

\begin{example}\label{e:Scope}
Consider $\scp(x_1)$, \texttt{Scope}~$\!(x_{1\hskip-0.07em}, \Fm)$, for $\Fm = (x_{1\hskip-0.12em} \X \nx3) \wedge (x_{1\hskip-0.12em} \X \nx2 \X x_3) \wedge (x_2 \X \nx3)$. $\scp(x_1) \gets \{x_1\}$ and $\Fm_* \hskip-0.12em\gets \Fm$ (L:1). Then, $\Fm_*^{\nx1}\hskip-0.15em$ is empty, and $\Fm_*^{x_1} \!= (x_{1\hskip-0.12em} \X \nx3) \wedge (x_{1\hskip-0.12em} \X \nx2 \X x_3)$ due to \texttt{OvrlEft}~$\!(x_{1\hskip-0.07em}, \Fm_*)$. Also, $\Ckk_*^{x_1} \!= \{{\color{red}1, 2}\}$, thus $c_{\color{red}1\hskip-0.05em} \gets \{x_3\}$ and $\lcl{*}(x_1) \gets \lcl{*}(x_1) \cup c_{\color{red}{1\hskip-0.07em}}$, as well as $c_{\color{red}2} \gets \{x_2, \nx3\}$ and $\lcl{*}(x_1) \gets \lcl{*}(x_1) \cup c_{\color{red}2\hskip-0.07em}$ (see \texttt{OvrlEft} L:1-6). Then, $\lcl{*}(x_1) = \{x_3, x_2, \nx3\}$ \& $\lcln_*(\neg\nx1) \gets \Fm_*^{\nx1}\hskip-0.15em$ (\texttt{OvrlEft} L:11). As a result, $\scp(x_1) \gets \scp(x_1) \cup \{x_1\} \cup \lcl{*}(x_1)$ (\texttt{Scope} L:4), and $ \scp(x_1) \supseteq \{x_3, \nx3\}$ (L:5), that is, $x_{1\hskip-0.09em} \imp x_{3\hskip-0.05em} \wedge \nx{3\hskip-0.05em}$, hence $x_{1\hskip-0.15em}$ is incompatible in the \emph{first} scan.
\end{example}

\begin{definition}\label{d:LFm}
$\Li^{_\scp\!\hskip-0.07em} = \{i\hskip-0.05em \in \Li \,|\, \lt_{i\hskip-0.12em} \in \scp_s\}$ and $\Li^{_\Fm\!} = \{i\hskip-0.05em \in \Li \,|\, \lt_{i\hskip-0.12em} \in C_{k\hskip-0.15em} \text{ in } \Fm_s\}$ due to $\FM_{s\hskip-0.11em} = \scp_{s\hskip-0.07em} \wedge \Fm_{s\hskip-0.05em}$.
\end{definition}

\texttt{Scan}~$\!(\FM_s)$ decomposes $\Fm_{s\hskip-0.11em}$ into $\scp_s(x_1), \scp_s(\nx1), \ldots, \scp_s(x_n), \scp_s(\nx n)$, \emph{whenever} $\Li^{_\scp\!\hskip-0.07em} \cap \Li^{_\Fm\!} = \emptyset$. If $\unsat \scp_{s - 1}(\lt_i)$, then $\nlt_{i\hskip-0.07em}$ is placed in $\scp_{s\hskip-0.05em}$, and leads to reductions of some $C_{k\hskip-0.12em}$ in $\Fm_{s\hskip-0.05em}$. In Figure \ref{f:cover}, $\unsat \scp_{s - 2}(\nx1)$ and $\unsat \scp_{s - 1}(x_3)$ hold, thus $\scp_{s\hskip-0.12em} = x_{1\hskip-0.12em} \wedge \nx{3\hskip-0.03em}$ and $\Fm_{s\hskip-0.12em} = (x_{4\hskip-0.09em} \X \nx{2\hskip-0.07em} \X x_n) \wedge \cdots \wedge (x_{2\hskip-0.07em} \X \nx n)$.

\begin{figure} [!h]
\vspace{-0.25em}
\begin{center}
  $\FM_{s\hskip-0.09em} = \underset{\displaystyle \scp_{s\hskip-0.07em}}{\underbrace{x_{1\hskip-0.12em} \wedge \nx{3\hskip-0.03em}}} \wedge
  \underset{\displaystyle \Fm_{s\hskip-0.07em}}{\underbrace{
        \underset{C_1}{\underbrace{
            (x_{4\hskip-0.09em} \X \nx{2\hskip-0.07em} \X x_n)
            }} \wedge \cdots \wedge
                \overset{\displaystyle \scp_s(\nx6) = \nx{6\hskip-0.07em} \wedge \nx{8\hskip-0.07em} \wedge x_{9\hskip-0.07em} \wedge \nx{4\hskip-0.07em} \wedge x_{7\hskip-0.07em}}{\overbracket[0.6pt]{
                (\nx{6\hskip-0.07em} \X x_8) \wedge (\nx{6\hskip-0.07em} \X \nx{9\hskip-0.07em} \X x_4) \wedge (x_{7\hskip-0.07em} \X x_8)
                }} \wedge \cdots \wedge
                    \underset{C_m}{\underbrace{(x_{2\hskip-0.07em} \X \nx n)}}
    }}$
\end{center}\vspace{-1.4em}
  \caption{\texttt{Scan}~$\!(\FM_s)$ decomposes $\Fm_{s\hskip-0.07em}$ into $\scp_s(x_1), \scp_s(\nx1), \ldots, \scp_s(x_n), \scp_s(\nx n)$, unless $\scp_s(.) \nsupseteq \{x_i, \nx i\}$}\label{f:cover}
\vspace{-0.25em}
\end{figure}

If $\nlt_{i\hskip-0.12em} \in \scp_{s\hskip-0.05em}$, then $\nlt_{i\hskip-0.12em}$ is necessary, thus $\lt_{i\hskip-0.12em}$ is incompatible \emph{trivially} for each $C_{k\hskip-0.09em} \ni \lt_{i\hskip-0.12em}$ in $\Fm_{s\hskip-0.07em}$ (see \texttt{Scan} L:1-2). For example, if ${\color{white!30!blue}x_{1\hskip-0.12em}} \wedge (x_{\color{red}1\hskip-0.10em} \X x_{2\hskip-0.05em} \X \nx3)$ holds, then $\nx{1\hskip-0.17em}$ becomes incompatible trivially. Note that ${\color{red}1} \hskip-0.09em\in \Li^{_\Fm\!}$ and ${\color{white!30!blue}x_{1\!}} \in \scp_{s\hskip-0.05em}$, and that $\nx{1\hskip-0.15em} \imp \nx{1\hskip-0.12em} \wedge x_{1\hskip-0.07em}$. If $\lt_{i\hskip-0.07em} \imp x_{j\hskip-0.09em} \wedge \nx{j\hskip-0.05em}$, then $\lt_{i\hskip-0.09em}$ is incompatible \emph{nontrivially} (L:6). See also Note \ref{n:incmp}/\ref{n:Incmp}. If \texttt{Scan}~$\!(\FM_s)$ is interrupted by \texttt{Remove} L:3, then $\FM$ is unsatisfiable. If it terminates (L:9), then a satisfying assignment is determined (Section \ref{s:DetAsg}).

\begin{note}\label{n:Incmp}
It is obvious that $\unsat \FM_s(\lt_j)$ if $\unsat (\scp_{s\hskip-0.05em} \wedge \lt_j)$ or $\unsat (\lt_{j\hskip-0.07em} \wedge \Fm_s)$ by Definition \ref{d:formula}/\ref{d:Fmri}, because $\FM_s(\lt_j) \hskip-0.07em= \scp_{s\hskip-0.05em} \wedge \lt_{j\hskip-0.07em} \wedge \Fm_{s\hskip-0.05em}$, and $\lt_{j\hskip-0.09em} \wedge \Fm_{s\hskip-0.09em} = \Fm_s(\lt_j)$, and that $\unsat \FM_s(\lt_j)$ iff $\neg \lt_{j\hskip-0.12em}$ holds (see Definition \ref{d:Incm}).
\end{note}

\begin{algorithm} \caption{\texttt{Scan}~$\!(\FM_s)$ \Cmmnt{$\FM_{s\hskip-0.11em} = \scp_{s\hskip-0.05em} \wedge \Fm_{s\hskip-0.05em}$, $\scp_{s\hskip-0.11em} = \hskip-0.15em\bigwedge\hskip-0.15em \lt_{i\hskip-0.12em}$ and $\Fm_{s\hskip-0.11em} = \hskip-0.11em\bigwedge\hskip-0.11em C_{k\hskip-0.05em}$. Checks if $\unsat \FM_s(\lt_i)$ for all $i\hskip-0.05em \in \Li^{_\Fm\!}$}}
\begin{algorithmic}[1]
\ForAll {$i \in \Li^{_\Fm\!}$ and $\nlt_{i\hskip-0.09em} \in \scp_{s\hskip-0.07em}$} \Cmmnt{Because $\nlt_{i\hskip-0.09em} \in \scp_{s\hskip-0.05em}$, $\unsat (\scp_{s\hskip-0.07em} \wedge \lt_{i\hskip-0.05em})$, that is, $\lt_{i\hskip-0.09em} \imp x_{i\hskip-0.09em} \wedge \nx{i\hskip-0.05em}$}
    \State \texttt{Remove}~$\!(\lt_i, \Fm_s)$; \Cmmnt{$\nlt_{i\hskip-0.12em}$ is \emph{necessary}, thus $\lt_{i\hskip-0.12em}$ is incompatible \emph{trivially}, hence $\nlt_{i\hskip-0.09em} \imp \neg \lt_{i\hskip-0.05em}$}
\EndFor \Cmmnt{If $i \in \Li^{_\scp}\!$, $\lt_{i\hskip-0.09em}$ has been already removed, hence $\nlt_{i\hskip-0.09em} \in \scp_{s\hskip-0.07em}$ and $\nlt_{i\hskip-0.09em} \notin C_{k\hskip-0.09em} \hskip0.1em \forall k \in \Ckk_{s\hskip-0.05em}$, i.e., $i \notin \Li^{_\Fm}\!$}
\ForAll {$i \in \Li^{_\Fm\!}$} \Cmmnt{$\Li^{_\scp\!} \cap \Li^{_\Fm\!} = \emptyset$ due to L:1-3. Hence, $i \in \Li^{_\scp\!\hskip-0.07em}$ iff $\lt_{i\hskip-0.05em} = x_{i\hskip-0.12em}$ is \emph{fixed} or $\lt_{i\hskip-0.05em} = \nx{i\hskip-0.12em}$ is \emph{fixed}}
    \ForAll {$\lt_{i\hskip-0.09em} \in \{x_i, \nx i\}$} \Cmmnt{Each and every $x_{i\hskip-0.09em}$ and $\nx{i\hskip-0.09em}$ \emph{assumed} compatible is to be \emph{verified}}
        \State \textbf{if} \texttt{Scope}~$\!(\lt_{i\hskip-0.05em}, \Fm_s)$ is NULL \textbf{then} \texttt{Remove}~$\!(\lt_{i\hskip-0.05em}, \Fm_s)$;  \Cmmnt{$\unsat \Fm_s(\lt_i)$, incompatible \emph{nontrivially}}
    \EndFor \Cmmnt{If $\lt_{i\hskip-0.12em} \imp x_{j\hskip-0.12em} \wedge \nx{j\hskip-0.05em}$, hence $\neg x_{j\hskip-0.12em} \vee \neg \nx{j\hskip-0.12em} \imp \neg \lt_{i\hskip-0.05em}$, then $\neg \lt_{i\hskip-0.12em} \imp \nlt_{i\hskip-0.05em}$, where $i \neq j$ due to L:1-3}
\EndFor \Cmmnt{$\neg \lt_{i\hskip-0.17em}$ iff $\nlt_{i\hskip-0.05em}$, since $\neg \lt_{i\hskip-0.11em} \imp \nlt_{i\hskip-0.10em}$ due to nontrivial, and $\neg \lt_{i\hskip-0.11em} \Leftarrow \nlt_{i\hskip-0.10em}$ due to trivial incompatibility}
\color{white!30!blue} \State \Return $\hat{\FM} \hskip-0.05em= \hat{\scp} \wedge \hat{\Fm}$, and $\scp(\lt_i) \:\&\: \Fm'(\lt_i)$ for all $i\hskip-0.07em \in \Li^{_{\hat{\Fm}}}$; \Cmmnt{$\hat{\scp} \gets \scp_{\hat{s}\hskip-0.07em}$ and $\hat{\Fm} \gets \Fm_{\hat{s}\hskip-0.05em}$. See also Note \ref{n:Trmnts}}
\end{algorithmic}
\label{alg:Scan}
\end{algorithm}

\begin{note}\label{n:Lprt}
$\Li^{_\scp\!\hskip-0.07em}$ and $\Li^{_\Fm\!}$ form a partition of $\Li$ due to Definition \ref{d:LFm} and \texttt{Scan} L:1-3.
\end{note}

\begin{note}\label{n:Trmnts}
When \texttt{Scan} terminates, $\hat{\scp}$ and $\hat{\Fm}$ become disjoint, and $\hat{\Fm} \equiv \hskip-0.05em\bigwedge_{i \in \Li}\hskip-0.11em \big(\scp(x_i) \oplus \scp(\nx i)\big)$, where $\Li \gets \Li^{_{\hat{\Fm}}}\hskip-0.12em$. Also, $\hat{\scp} = \hskip-0.11em\bigwedge\hskip-0.11em \lt_{i\hskip-0.09em}$ and $\hat{\Fm} = \hskip-0.11em\bigwedge\hskip-0.11em C_{k\hskip-0.09em}$ such that $|C_k| > 1$, because each $C_{k\hskip-0.12em} = \{\lt_i\}$ in $\Fm_{s\hskip-0.07em}$ for any $s$ transforms into $\lt_{i\hskip-0.12em}$ in $\hat{\scp}$. That is, $C_{k\hskip-0.09em} = (\lt_{i\hskip-0.09em} \X \lt_j)$ or $C_{k\hskip-0.09em} = (\lt_{i\hskip-0.09em} \X \lt_{j\hskip-0.09em} \X \lt_u)$ in $\hat{\Fm}$.
\end{note}

\texttt{Remove}~$\!(\lt_j, \Fm_s)$ leads to reductions of any $C_{k\hskip-0.09em} \ni \nlt_{j\hskip-0.12em}$ due to $\nlt_{j\hskip-0.05em}$, which consists in $\scp_{s+1\hskip-0.07em}$ (see L:1-2), as well as of any $C_{k\hskip-0.09em} \ni \lt_{j\hskip-0.09em}$ due to $\neg \lt_{j\hskip-0.05em}$, which consists in $\Fm_{s+1\hskip-0.12em}$ (see L:1,5).

\begin{algorithm}\caption{\texttt{Remove}~$\!(\lt_j, \Fm_s)$\Cmmnt{$\lt_{j\hskip-0.12em}$ is incompatible/removed iff $\nlt_{j\hskip-0.12em}$ is necessary, i.e., $\neg \lt_{j\hskip-0.12em}$ iff $\nlt_{j\hskip-0.05em}$}}
\begin{algorithmic}[1]
\State \texttt{OvrlEft}~$\!(\nlt_j, \Fm_s)$; \Cmmnt{\texttt{OvrlEft} is defined over $\Fm_{s\hskip-0.11em} = \hskip-0.11em\bigwedge\hskip-0.11em C_{k\hskip-0.05em}$, $|C_k| > 1$, and returns $\lcl{s}(\nlt_j)$ \& $\lcln_s(\neg \lt_j)$}
\State $\scp_{s+1\hskip-0.12em} \gets \scp_{s\hskip-0.07em} \cup \{\nlt_j\} \cup \lcl{s}(\nlt_j)$; \Cmmnt{$\scp_{s+1\hskip-0.11em} = \hskip-0.11em\bigwedge\hskip-0.11em \lt_{i\hskip-0.12em}$ is true by definition, unless $\scp_{s+1\hskip-0.07em}$ involves $x_{i\hskip-0.08em} \wedge \nx{i\hskip-0.05em}$}
{\color{red} \State \textbf{if} $\scp_{s+1\hskip-0.07em} \supseteq \{x_i, \nx i\}$ for some $i$ \textbf{then} \Return $\FM$ is unsatisfiable; \Cmmnt{$\FM_{s\hskip-0.11em} = \scp_{s\hskip-0.07em} \wedge \Fm_{s\hskip-0.07em}$}}
\State $\Li^{_\Fm\!} \gets \Li^{_\Fm\!} - \{j\}$; $\Li^{_\scp\!\hskip-0.07em} \gets \Li^{_\scp\!} \cup \{j\}$;
\State $\Fm_{s+1} \hskip-0.15em \gets \lcln_s(\neg \lt_j) \wedge \Fm'_{s\hskip-0.05em}$; Update $\{C_k\}$ over $\Fm_{s+1\hskip-0.05em}$; \Cmmnt{\hskip-0.05em$\Fm'_{s\hskip-0.12em}$ denotes clauses beyond the entire $\scp_{s\hskip-0.07em}$ effect} \linebreak
  \Cmmnt{$\Fm'_{s\hskip-0.11em} = \hskip-0.11em\bigwedge\hskip-0.11em C_{k\hskip-0.09em}$ for $k \in \Ckk'_s$, where $\Ckk'_{s\hskip-0.11em} = \Ckk_s - (\Ckk_s^{\nx j\hskip-0.15em} \cup \Ckk_s^{x_j})$, and $\Ckk_s^{\nx j\hskip-0.15em} \cap \Ckk_s^{x_j\hskip-0.09em} = \emptyset$ due to Lemma \ref{l:Cnvrt}}
\color{white!30!blue} \State \texttt{Scan}~$\!(\FM_{s+1})$; \Cmmnt{$\lt_{i\hskip-0.12em}$ verified compatible for $\check{s} \hskip-0.03em \leqslant s$ can be incompatible for $\tilde{s} \hskip-0.03em > s$ due to $\neg \lt_{j\hskip-0.12em}$ in $\Fm_s$}
\end{algorithmic}
\label{alg:Remove}
\end{algorithm}
\subsection{\texorpdfstring{Satisfiability of the Formula $\FM$ vs Satisfiability of the Scope $\scp(\lt_i)$}{Satisfiability}}\label{s:UnSat}
This section shows that $\FM$ is satisfiable iff $\scp(\lt_i)$ is satisfied for all $i \in \Li$, and any $\lt_{i\hskip-0.09em} \in \{x_i, \nx{i}\}$.

\begin{proposition}[Nontrivial incompatibility]\label{p:Incmp}
$\unsat \Fm_s(\lt_j)$ iff $\unsat \scp_s(\lt_j)$ or $\unsat \Fm'_s(\lt_j)$ for any $s$.
\end{proposition}
\begin{proof}
Proof is obvious due to $\Fm_s(\lt_j) = \scp_s(\lt_j) \wedge \Fm'_s(\lt_j)$ by Lemma \ref{l:Scope}.
\end{proof}

\begin{note}[Assumption]\label{n:assmp}
$\unsat \Fm_s(\lt_j)$ is verified \emph{solely} via $\unsat \scp_s(\lt_j)$ for some $s$, whether or not $\unsat \Fm'_s(\lt_j)$ is \emph{ignored}, which is sufficient for incompatibility, and easy to check (see \texttt{Scope} \hyperref[alg:Scope]{L:5}).
\end{note}

The following introduces the tools to justify this assumption, which facilitates the $\FM$ scan. Assume that \texttt{Scan} \emph{terminates} (\hyperref[alg:Scan]{L:9}), that is, $\scp \wedge \Fm$ transforms into $\hat{\scp} \wedge \hat{\Fm}$. Let $\Fm \gets \hat{\Fm}$, thus $\Li \gets \Li^{_{\hat{\Fm}\!}}$. Therefore, $\lt_{i\hskip-0.12em} \ent \scp(\lt_i)$ for all $i \in \Li$ and $\lt_{i\hskip-0.12em} \in \{x_i, \nx{i}\}$. That is, as $\lt_{i\hskip-0.12em} = \mathbf{T}$, $\scp(\lt_i) = \mathbf{T}$.

\begin{definition}
$\Li(.) \hskip-0.07em= \Li(\scp(.))$ and $\Li'(.) \hskip-0.07em= \Li(\Fm'(.))$, which denote respective index sets.
\end{definition}

\begin{lemma}[No conjunct exists in beyond the scope]\label{l:Indp}
$\Li(\lt_j) \cap \Li'(\lt_j) = \emptyset$ for any $j \hskip-0.03em\in \Li$.
\end{lemma}
\begin{proof}
$\Fm'(\lt_j) \hskip-0.05em= \hskip-0.11em\bigwedge\hskip-0.11em C_{k\hskip-0.09em}$ due to Lemma \ref{l:Scope}. Let $\lt_{i\hskip-0.12em}$ the \emph{conjunct} be in $C_{k\hskip-0.05em}$, i.e., $i \hskip-0.03em\in \big(\Li(\lt_j) \cap \Li'(\lt_j)\big)$. Then, for any $C_{k\hskip-0.09em} \ni \lt_{i\hskip-0.05em}$, $(\lt_{i\hskip-0.09em} \X x_{j\hskip-0.09em} \X \nx u) \clp (\lt_{i\hskip-0.07em} \wedge \nx{j\hskip-0.09em} \wedge x_u)$, thus $\lt_{i\hskip-0.09em} \notin C_{k\hskip-0.05em}$. Moreover, for any $C_{k\hskip-0.09em} \ni \nlt_{i\hskip-0.05em}$, $(\nlt_{i\hskip-0.09em} \X \lt_{v\hskip-0.09em} \X \lt_y) \rdc (\lt_{v\hskip-0.09em} \X \lt_y)$, thus $\nlt_{i\hskip-0.09em} \notin C_{k\hskip-0.05em}$. See Definition \ref{d:colp}/\ref{d:shrn}. Hence, $i \notin \big(\Li(\lt_j) \cap \Li'(\lt_j)\big)$.
\end{proof}

$\scp(\lt_i | \lt_j)$ is called the conditional scope, and $\Fm'(\lt_i | \lt_j)$ is called conditional beyond the scope, which are defined over $\Fm'(\lt_j)$ for $j \neq i$, that is, constructed by \texttt{Scope}~$\!\big(\lt_i, \Fm'(\lt_j)\big)$.

\begin{lemma}\label{l:prtn}
$\Li$ is partitioned into $\Li(\lt_j), \,\Li(\lt_{j_1\hskip-0.03em} | \lt_j), \,\Li(\lt_{j_2} | \lt_{j_1}), \ldots, \Li(\lt_{j_n} | \lt_{j_m})$, thus $\Fm(\lt_j)$ is decomposed into \emph{disjoint} $\scp(\lt_j), \,\scp(\lt_{j_1\hskip-0.03em} | \lt_j), \,\scp(\lt_{j_2} | \lt_{j_1}), \ldots, \scp(\lt_{j_n} | \lt_{j_m})$.
\end{lemma}
\begin{proof}
\texttt{Scope}~$\!(\lt_j, \Fm)$ \emph{partitions} $\Li$ into $\Li(\lt_j)$ and $\Li'(\lt_j)$ for any $j \hskip-0.05em\in \Li$ (see also Lemma \ref{l:Indp}). Thus, $\Fm(\lt_j)$ is decomposed into \emph{disjoint} $\scp(\lt_j)$ and $\Fm'(\lt_j)$. Then, \texttt{Scope}~$\!\big(\lt_{j_1\hskip-0.09em}, \Fm'(\lt_j)\big)$ \emph{partitions} $\Li'(\lt_j)$ into $\Li(\lt_{j_1\hskip-0.03em} | \lt_j)$ and $\Li'(\lt_{j_1\hskip-0.03em} | \lt_j)$ for any $j_1 \hskip-0.09em\in \Li'(\lt_j)$. Thus, $\Fm'(\lt_j)$ is decomposed into \emph{disjoint} $\scp(\lt_{j_1\hskip-0.03em} | \lt_j)$ and $\Fm'(\lt_{j_1\hskip-0.03em} | \lt_j)$. Finally, $\Fm'(\lt_{j_m} | \lt_{j_l})$ is decomposed into \emph{disjoint} $\scp(\lt_{j_n} | \lt_{j_m})$ and $\Fm'(\lt_{j_n} | \lt_{j_m})$ for any $j_{n\hskip-0.09em} \in \Li'(\lt_{j_m} | \lt_{j_l})$ such that $\Li'(\lt_{j_n} | \lt_{j_m}) = \emptyset$ (see also Note \ref{n:ScpTrmnts}).
\end{proof}

\begin{lemma}\label{l:Prtn}
$\Fm'(\lt_j)$ is decomposed into \emph{disjoint} $\scp(\lt_{j_1\hskip-0.03em} | \lt_j), \, \scp(\lt_{j_2} | \lt_{j_1}), \ldots, \scp(\lt_{j_n} | \lt_{j_m})$.
\end{lemma}
\begin{proof}
Follows directly from Lemma \ref{l:prtn}, and from Lemma \ref{l:Scope}, $\Fm(\lt_j) = \scp(\lt_j) \wedge \Fm'(\lt_j)$.
\end{proof}

\begin{lemma}\label{l:SubSet}
$\Fm \supseteq \Fm'(\lt_j) \supseteq \Fm'(\lt_{j_1\hskip-0.03em} | \lt_j) \supseteq \Fm'(\lt_{j_2} | \lt_{j_1}) \supseteq \cdots \supseteq \Fm'(\lt_{j_m} | \lt_{j_l})$, when it \emph{terminates}.
\end{lemma}
\begin{proof}
Follows directly from Lemma \ref{l:prtn}. Then, some $C_{k\hskip-0.09em}$ in $\Fm$ collapse to some $c_{k\hskip-0.09em}$ in $\scp(\lt_j)$. Thus, the number of $C_{k\hskip-0.09em}$ in $\Fm$ is greater than or equal to that of $C_{k\hskip-0.09em}$ in $\Fm'(\lt_j)$, hence $|\Ckk| \GE |\Ckk'|$, where $\Ckk$ is an index set of $C_{k\hskip-0.09em}$ in $\Fm$. Also, some $C_{k\hskip-0.09em}$ in $\Fm$ shrink to some $C_{k'}\hskip-0.12em$ in $\Fm'(\lt_j)$, hence $\forall k' \!\in \Ckk' \hskip0.1em \exists k \in \Ckk \, [C_{k\hskip-0.12em} \supseteq C_{k'}]$. Thus, $\Fm \supseteq \Fm'(\lt_j)$. Likewise, $\Fm'(\lt_j) \supseteq \Fm'(\lt_{j_1\hskip-0.03em} | \lt_j)$, because $\Fm'(\lt_j)$ is decomposed into $\scp(\lt_{j_1\hskip-0.03em} | \lt_j)$ and $\Fm'(\lt_{j_1\hskip-0.03em} | \lt_j)$. Therefore, $\Fm \supseteq \Fm'(\lt_j) \supseteq \Fm'(\lt_{j_1\hskip-0.03em} | \lt_j) \supseteq \Fm'(\lt_{j_2} | \lt_{j_1}) \supseteq \cdots \supseteq \Fm'(\lt_{j_m} | \lt_{j_l})$, where $\Fm'(\lt_{j_m} | \lt_{j_l}) = \Fm'(\lt_{j_m} | \lt_j, \ldots, \lt_{j_l})$. Note that $\Fm'(\lt_{j_n} | \lt_{j_m}) = \big\{\{\}\big\}$.
\end{proof}

\begin{lemma}\label{l:Ent}
$\scp(\lt_i) \ent \scp(\lt_i | \lt_j)$, thus $\scp(\lt_i) \prv \scp(\lt_i | \lt_j)$, when the scan \emph{terminates}.
\end{lemma}
\begin{proof}
\texttt{Scope}~$\!(\lt_i, \Fm)$ constructs $\scp(\lt_i)$ and \texttt{Scope}~$\!\big(\lt_i, \Fm'(\lt_j)\big)$ constructs $\scp(\lt_i | \lt_j)$. $\Fm \supseteq \Fm'(\lt_j)$ by Lemma \ref{l:SubSet}. Therefore, $\scp(\lt_i) \supseteq \scp(\lt_i | \lt_j)$, and $\scp(\lt_i) \ent \scp(\lt_i | \lt_j)$ (see also Figure \ref{f:OvCntn}), where $\scp(\lt_i) = \lt_{i\hskip-0.07em} \wedge \lt_{j\hskip-0.09em} \wedge \cdots \wedge \lt_{v\hskip-0.09em}$ and $\scp(\lt_i | \lt_j) = \lt_{i\hskip-0.05em} \wedge \cdots \wedge \lt_{v\hskip-0.05em}$. Then, $\lt_{j\hskip-0.09em} \notin \scp(\lt_i | \lt_j)$, since $\lt_{j\hskip-0.09em} \notin C_{k\hskip-0.09em}$ for any $C_{k\hskip-0.09em} \in \Fm'(\lt_j)$ by Lemma \ref{l:Indp}. It is obvious that if $\scp(\lt_i) \ent \scp(\lt_i | \lt_j)$, then $\scp(\lt_i) \prv \scp(\lt_i | \lt_j)$.
\end{proof}

Lemma \ref{l:Ent} leads to Lemma \ref{l:EntGn}, because $\lt_{i\hskip-0.12em} \ent \scp(\lt_i)$ and $\lt_{i\hskip-0.12em} \prv \scp(\lt_i)$ by Lemma \ref{l:Scope}. That is, each and every conditional scope $\scp(\lt_i | .)$ is entailed and proved, when the scan \emph{terminates}.  

\begin{lemma}\label{l:EntGn}
$\scp(\lt_i | \lt_j), \, \scp(\lt_i | \lt_j, \lt_{j_1}), \ldots, \scp(\lt_i | \lt_j, \lt_{j_1\hskip-0.09em}, \ldots, \lt_{j_m})$ holds for every $j \in \Li$, and for every $i \in \Li'(\lt_j)$, $i \in \Li'(\lt_{j_1\hskip-0.03em} | \lt_j), \ldots, i \in \Li'(\lt_{j_m} | \lt_j, \lt_{j_1\hskip-0.09em}, \ldots, \lt_{j_l})$, when the scan \emph{terminates}.
\end{lemma}
\begin{proof}
$\Fm \supseteq \Fm'(\lt_j) \supseteq \Fm'(\lt_{j_1\hskip-0.03em} | \lt_j) \supseteq \cdots \supseteq \Fm'(\lt_{j_m} | \lt_{j_l})$ by Lemma \ref{l:SubSet}. Hence, $\scp(\lt_i) \supseteq \scp(\lt_i | \lt_j), \linebreak \scp(\lt_i) \supseteq \scp(\lt_i | \lt_j, \lt_{j_1}), \ldots, \scp(\lt_i) \supseteq \scp(\lt_i | \lt_j, \ldots, \lt_{j_m})$, and $\scp(\lt_i) \ent \scp(\lt_i | \lt_j), \scp(\lt_i) \ent \scp(\lt_i | \lt_j, \lt_{j_1}), \linebreak \ldots, \scp(\lt_i) \ent \scp(\lt_i | \lt_j, \lt_{j_1}, \ldots, \lt_{j_m})$. Note that if $\scp(\lt_i) \ent \scp(\lt_i | .)$, then $\scp(\lt_i) \prv \scp(\lt_i | .)$. Therefore, $\scp(\lt_i | \lt_j)$, $\scp(\lt_i | \lt_j, \lt_{j_1}), \ldots, \scp(\lt_i | \lt_j, \lt_{j_1\hskip-0.09em}, \ldots, \lt_{j_m})$ hold, which generalizes Lemma \ref{l:Ent}.
\end{proof}

\begin{theorem}[Unsatisfiability]\label{t:InCmp}
$\lt_{j\hskip-0.12em}$ is incompatible due to $\unsat \Fm(\lt_j)$ iff $\unsat \scp_s(\lt_j)$ for some $s$.
\end{theorem}

\begin{corollary}[Satisfiability]\label{c:Sat}
$\sat \Fm$ iff the scope $\scp(\lt_i)$ holds for every $i \in \Li$ and $\lt_{i\hskip-0.09em} \in \{x_i, \nx{i}\}$.
\end{corollary}
\begin{proof}
$\scp(\lt_{j_1\hskip-0.09em} | \lt_j)$, $\scp(\lt_{j_2} | \lt_{j_1}), \ldots, \scp(\lt_{j_n} | \lt_{j_m})$ defined over $\Fm'(\lt_j)$ are disjoint due to Lemma \ref{l:Prtn} such that $\scp(\lt_{j_1\hskip-0.09em} | \lt_j)$, $\scp(\lt_{j_2} | \lt_{j_1}), \ldots, \scp(\lt_{j_n} | \lt_{j_m})$ hold by Lemma \ref{l:EntGn} for any $j \in \Li$, $j_{1\hskip-0.12em} \in \Li'(\lt_j)$, $j_{2\hskip-0.05em} \in \Li'(\lt_{j_1\hskip-0.09em} | \lt_j), \ldots, {j_n\hskip-0.05em} \in \Li'(\lt_{j_m} | \lt_{j_l})$, thus $\Fm'(\lt_j)$ is composed of $\scp(.)$ both disjoint and satisfied. Therefore, $\Fm'(\lt_j)$ is \emph{satisfiable}, and unsatisfiability of $\Fm'_s(\lt_j)$ is \emph{ignored} to verify $\unsat \Fm_s(\lt_j)$. Hence, Theorem \ref{t:InCmp} holds (see Proposition~\ref{p:Incmp} and Note \ref{n:assmp}). Then, $\scp(\lt_i) \equiv \Fm(\lt_i)$, since $\Fm'(\lt_i)$ is satisfiable, and $\Fm(\lt_i) \hskip-0.05em= \scp(\lt_i) \wedge \Fm'(\lt_i)$. Thus, Corollary \ref{c:Sat} holds (see also Appendix \ref{s:App}).
\end{proof}

Theorem \ref{t:mntn} shows that any $\lt_{j\hskip-0.12em}$ incompatible remains incompatible, even if $\lt_{i\hskip-0.12em}$ is removed.

\begin{theorem}\label{t:mntn}
If $\unsat \FM_{\tilde{s}}(\lt_j)$ for some $\tilde{s}$, then $\unsat \FM_s(\lt_j)$ for all $s > \tilde{s}$, even if $\neg \lt_{i\hskip-0.09em}$ holds, $i \neq j$.
\end{theorem}
\begin{proof}
See Note \ref{n:Incmp}/\ref{n:Lprt}. $\unsat \FM_s(\lt_j)$ iff $\unsat (\scp_{s\hskip-0.05em} \wedge \lt_j)$ or $\unsat \Fm_s(\lt_j)$. Let $\unsat (\scp_{\tilde{s}\hskip-0.05em} \wedge \lt_j)$ for some $\tilde{s}$. Then, $\unsat (\scp_{s\hskip-0.05em} \wedge \lt_j)$ for all $s > \tilde{s}$, since $\scp_{\tilde{s}\hskip-0.07em} \subseteq \scp_{s\hskip-0.09em}$ due to \texttt{Remove} \hyperref[alg:Remove]{L:2}. Let $\unsat \Fm_{\tilde{s}}(\lt_j)$ due to \emph{solely} $x_{i\hskip-0.07em} \wedge \nx{i\hskip-0.05em}$. Then, $\color{red}\nx{i\hskip-0.11em} \vee x_{i\hskip-0.09em} \imp \nlt_{j\hskip-0.05em},$ thus $\nlt_{j\hskip-0.09em} \in \scp_{s\hskip-0.07em}$ for $s > \tilde{s}$. Hence, $\unsat (\scp_{s\hskip-0.05em} \wedge \lt_j)$ for all $s > \tilde{s}$. Assume that $\lt_{i\hskip-0.12em}$ is removed \emph{before} $\lt_{j\hskip-0.05em}$, that is, $\neg \lt_{i\hskip-0.09em}$ holds by $\unsat \FM_{\check{s}}(\lt_i)$ for $\check{s} \LE \tilde{s}$. Then, $\neg \lt_{i\hskip-0.09em} \imp \nlt_{i\hskip-0.09em}$ and $\color{red}\nlt_{i\hskip-0.09em} \imp \nlt_{j\hskip-0.05em},$ thus $\{\nlt_i, \nlt_j\} \subseteq \scp_{s\hskip-0.07em}$ for $s > \tilde{s}$. Note that $\scp_{\check{s}\hskip-0.07em} \subseteq \scp_{\tilde{s}\hskip-0.07em} \subseteq \scp_{s\hskip-0.05em}$. Hence, $\unsat (\scp_{s\hskip-0.05em} \wedge \lt_{i\hskip-0.07em} \wedge \lt_j)$ for all $s > \tilde{s}$. If $\lt_{i\hskip-0.12em}$ is removed \emph{after} $\lt_{j\hskip-0.05em}$, i.e., $\neg \lt_{i\hskip-0.09em}$ holds by $\unsat \FM_s(\lt_i)$ for $s > \tilde{s}$, then $\unsat (\scp_{s\hskip-0.05em} \wedge \lt_{j\hskip-0.07em} \wedge \lt_i)$ for all $s > \tilde{s}$.
\end{proof}

\begin{proposition}
The time complexity of \texttt{\emph{Scan}} is $O(mn^3)$.
\end{proposition}
\begin{proof}
\hyperref[alg:Ovrl]{\texttt{OvrlEft}}, and \hyperref[alg:Remove]{\texttt{Remove}}, takes $4m$ steps by $\big(|\Ckk_*^{\lt_j}| \times |C_k|\big) + |\Ckk_*^{\nlt_j}| = 3m + m$. \hyperref[alg:Scope]{\texttt{Scope}} takes $n4m$ steps by $|\scp_s(\lt_j)| \times 4m$. Then, \hyperref[alg:Scan]{\texttt{Scan}} takes $n^24m$ steps due to L:1-3 by $|\Li^{_\Fm}| \times |\scp_s| \times 4m$, as well as $8n^2m + 8nm$ steps due to L:4-8 by $2|\Li^{_\Fm}| \times (4nm + 4m)$. Also, the number of the scans is $\hat{s} \LE |\Li^{_\Fm}|$ due to \texttt{Remove} \hyperref[alg:Remove]{L:6}. Therefore, the time complexity of \texttt{Scan} is $O(n^3m)$.
\end{proof}

\begin{example}\label{e:example2}
${\color{purple}\FM} = \big\{\{\}, \{x_3, x_4, \nx5\}, \{x_3, x_6, \nx7\}, \{x_4, x_6, \nx7\}\big\}$, i.e., $\scp = \emptyset$. Let \texttt{Scope}~$\!(x_3, \Fm)$ execute \emph{first} in the \emph{first} scan, which leads to the reductions below over $\Fm$ due to $x_3$.

\begin{tabular}{  r  m{0mm}  m{2cm}  m{0mm}  m{2cm}  m{0mm}  m{2cm}  m{3cm} }
  $\Fm(x_3)$ & $\hskip-1em=$    & $\hskip-1em\Co$  & $\hskip-2em\wedge$ & $\hskip-2.4em\Ct$  & $\hskip-3.4em\wedge$ & $\hskip-3.8em\CT$& $\hskip-4.7em\wedge\:{\color{white!30!blue}x_3}$\\
      $x_3$  & $\hskip-1em\imp$ & $\hskip-1em\Cfb$ & $\hskip-2em\wedge$ & $\hskip-2.4em\CF$  & $\hskip-3.4em\wedge$ & $\hskip-3.8em\CT$& $\hskip-4.7em\wedge\: x_3$\\
      $\nx4$ & $\hskip-1em\imp$ & $\hskip-1em\Cf$  & $\hskip-2em\wedge$ & $\hskip-2.4em\CFb$ & $\hskip-3.4em\wedge$ & $\hskip-3.8em\Cs$& $\hskip-4.7em\wedge\: x_3$\\
      $\nx6$ & $\hskip-1em\imp$ & $\hskip-1em\Cf$  & $\hskip-2em\wedge$ & $\hskip-2.4em\CF$  & $\hskip-3.4em\wedge$ & $\hskip-3.8em\CS$& $\hskip-4.7em\wedge\: x_3$
\end{tabular}

Since $\unsat \big(\scp(x_3) = x_3 \wedge \nx4 \wedge x_5 \wedge \nx6 \wedge x_7 \wedge \nx7\big)$, $x_3$ is incompatible, hence $\neg x_3 \imp \nx3$, that is, $\nx3$ is necessary. Thus, $\FM \Rdc \FM_2$ by $(x_3 \X x_{4\hskip-0.05em} \X \nx5) \rdc (x_{4\hskip-0.05em} \X \nx5)$ and $(x_3 \X x_6 \X \nx7) \rdc (x_6 \X \nx7)$. As a result, ${\color{purple}\FM_2} = \nx3 \wedge (x_{4\hskip-0.05em} \X \nx5) \wedge (x_6 \X \nx7)\wedge (x_{4\hskip-0.05em} \X x_6 \X \nx7)$. Let \texttt{Scope}~$\!(x_5, \Fm_2)$ execute next.

\begin{tabular}{  r  m{0mm}  m{2cm}  m{0mm}  m{2cm}  m{0mm}  m{2cm}  m{0mm} m{2cm} }
  $\Fm_2(x_5)$ & $\hskip-1em=$    & $\hskip-1em\co$ & $\hskip-2.4em\wedge$ & $\hskip-2.8em\ct$ & $\hskip-4.2em\wedge$ & $\hskip-4.6em\cT$ & $\hskip-5.5em\wedge\; x_5$\\
         $x_5$ & $\hskip-1em\imp$ & $\hskip-1em\cf$ & $\hskip-2.4em\wedge$ & $\hskip-2.8em\ct$ & $\hskip-4.2em\wedge$ & $\hskip-4.6em\cT$ & $\hskip-5.5em\wedge\; x_5$\\
         $x_4$ & $\hskip-1em\imp$ & $\hskip-1em\cf$ & $\hskip-2.4em\wedge$ & $\hskip-2.8em\ct$ & $\hskip-4.2em\wedge$ & $\hskip-4.6em\cs$ & $\hskip-5.5em\wedge\; x_5$\\
        $\nx6$ & $\hskip-1em\imp$ & $\hskip-1em\cf$ & $\hskip-2.4em\wedge$ & $\hskip-2.8em\cF$ & $\hskip-4.2em\wedge$ & $\hskip-4.6em\cs$ & $\hskip-5.5em\wedge\; x_5$
\end{tabular}

Since $\unsat \big(\scp_2(x_5) = x_4 \wedge \nx7 \wedge \nx6 \wedge x_7 \wedge \nx3 \wedge x_5\big)$, $x_5$ is incompatible, hence $\neg x_5 \imp \nx5$. Thus, $\FM_2 \hskip-0.05em\Rdc \FM_3$ by $(x_{4\hskip-0.05em} \X \nx5) \clp (\nx{4\hskip-0.05em} \wedge \nx5)$, where ${\color{purple}\FM_3} = \nx3 \wedge \nx{4\hskip-0.05em} \wedge \nx5 \wedge (x_6 \X \nx7) \wedge (x_{4\hskip-0.05em} \X x_6 \X \nx7)$. Then, $\nx4\hskip-0.07em$ leads to the next reduction by $(x_{4\hskip-0.05em} \X x_6 \X \nx7) \rdc (x_6 \X \nx7)$, and \texttt{Scan}~$\!(\FM_4)$ \emph{terminates}. That is, $\hat{\FM} = \hat{\scp} \wedge \hat{\Fm}$, where $\hat{\scp} = \{\nx3, \nx4, \nx5\}$ and $\hat{\Fm} = \big\{\{x_6, \nx7\}\big\}$, since ${\color{purple}\FM_{4\hskip-0.09em}} = \nx3 \wedge \nx4 \wedge \nx5 \wedge (x_6 \X \nx7)$.

\end{example}

In Example \ref{e:example2}, if \texttt{Scope}~$\!(x_5, \Fm)$ executes \emph{first}, then $\scp(x_5) = x_{5\hskip-0.07em}$ becomes the scope, and $\Fm'(x_5) = (x_3 \X x_4) \wedge (x_3 \X x_6 \X \nx7) \wedge (x_4 \X x_6 \X \nx7)$ becomes beyond the scope of $x_{5\hskip-0.07em}$ over $\Fm$. Then, $x_{5\hskip-0.07em}$ is compatible (in $\Fm$) due to Theorem~\ref{t:InCmp}, since $\scp(x_5)$ holds, while it is incompatible due to Proposition~\ref{p:Incmp}, since $\unsat \Fm'(x_5)$ holds. On the other hand, the fact that $\unsat \Fm'(x_5)$ holds is verified indirectly. That is, incompatibility of $x_{5\hskip-0.07em}$ is checked by means of $\scp_s(x_5)$ for some $s$. Then, $x_{5\hskip-0.07em}$ becomes incompatible (in $\Fm_{\color{red}2}$), because $\unsat \scp_{\color{red}2}(x_5)$ holds, after $\FM \Rdc \FM_2$ by removing $x_{3\hskip-0.07em}$ from $\Fm$ due to $\unsat \scp(x_3)$. As a result, $\unsat \Fm'(x_5)$ holds due to $\neg x_3$. Thus, there exists no $\lt_{j\hskip-0.12em}$ such that $\unsat \Fm'(\lt_j)$, when the scan \emph{terminates}, because $\scp(\lt_i)$ holds for all $\lt_{i\hskip-0.12em}$ in $\Fm$, hence $\scp(\lt_i | \lt_j)$ holds for all $\lt_{i\hskip-0.12em}$ in $\Fm'(\lt_j)$, after each $\lt_{j\hskip-0.12em}$ is removed if $\unsat \scp_s(\lt_j)$ (see also Figures \ref{f:Assmpt}-\ref{f:cover}).
\subsection[Finding an assignment]{Construction of a satisfying assignment by composing scopes}\label{s:DetAsg}
$\hat{\FM} \hskip-0.05em= \hat{\scp} \wedge \hat{\Fm}$, when \texttt{Scan}~$\!(\FM_{\hat{s}})$ terminates. Let $\scp \hskip-0.05em\coloneqq \hat{\scp}$ and $\Fm \hskip-0.05em\coloneqq \hat{\Fm}$, i.e., $\Li \hskip-0.05em\coloneqq \Li^{_{\hat{\Fm}\!}}$. Then, $\sat \Fm$ holds by Corollary \ref{c:Sat}, where $\alpha$ is a satisfying assignment, and constructed by Algorithm \ref{alg:CnstAssg} through any $(i_0, i_{1\hskip-0.07em}, i_2, \ldots, i_m, i_n)$ over $\Li$ such that $\alpha = \{\scp(\lt_{i_0}), \scp(\lt_{i_1\hskip-0.05em} | \lt_{i_0}), \scp(\lt_{i_2} | \lt_{i_1\hskip-0.05em}), \ldots, \scp(\lt_{i_n} | \lt_{i_m})\}$. Thus, $\FM$ is decomposed into \emph{disjoint} scopes $\scp, \scp(\lt_{i_0}), \scp(\lt_{i_1\hskip-0.05em} | \lt_{i_0}), \scp(\lt_{i_2} | \lt_{i_1}), \ldots, \scp(\lt_{i_n} | \lt_{i_m})$ (see Note \ref{n:Lprt}, and Lemma \ref{l:prtn}). Recall that any scope $\scp(.)$ denotes a minterm by Definition \ref{d:minterm}/\ref{d:formula}, and that \texttt{Scope}~$\!(\lt_i, \Fm)$ constructs $\scp(\lt_i)$ and $\Fm'(\lt_i)$ to determine a satisfying assignment, unless $\FM$ collapses to a \emph{unique} assignment, that is, unless $\hat{\FM} = \alpha = \hat{\scp}$. See also Appendix \ref{s:App} to determine a satisfying assignment without constructing $\scp(\lt_i | .)$ by \texttt{Scope}~$\!\big(\lt_i, \Fm'(.)\big)$.

\begin{algorithm}\caption{\Cmmnt{Construction of a satisfying assignment $\alpha$ over $\Fm$, $\Li \coloneqq \Li^{_{\hat{\Fm}}\!}$ and $\Fm \coloneqq \hat{\Fm}$}}
\begin{algorithmic}
    \State Pick $j \in \Li$; \Cmmnt{The scope $\scp(\lt_i)$ and beyond the scope $\Fm'(\lt_i)$ for all $i \in \Li$ are available initially}
    \State $\alpha \gets \scp(\lt_j)$; $\Li \gets \Li - \Li(\lt_j)$; $\Fm \gets \Fm'(\lt_j)$;
    \Repeat
        \State Pick $i \in \Li$; \texttt{Scope}~$\!(\lt_i, \Fm)$; \Cmmnt{It constructs $\scp(\lt_i | \lt_j)$ and $\Fm'(\lt_i | \lt_j)$ with respect to $\Fm'(\lt_j)$}
        \State $\alpha \gets \alpha \cup \scp(\lt_i)$; \Cmmnt{$\scp(\lt_i) \coloneqq \scp(\lt_i | \lt_j)$, because $\scp(\lt_i)$ is \emph{unconditional} with respect to $\Fm$ updated}
        \State $\Li \gets \Li - \Li(\lt_i)$; \Cmmnt{$\Li \gets \Li'(\lt_i | \lt_j)$ due to the partition $\big\{\Li(\lt_j), \Li(\lt_i | \lt_j), \Li'(\lt_i | \lt_j)\big\}$ over $\Li$}
        \State $\Fm \gets \Fm'(\lt_i)$; \Cmmnt{$\Fm'(\lt_i) \coloneqq \Fm'(\lt_i | \lt_j)$, because $\Fm'(\lt_i)$ is \emph{unconditional} with respect to $\Fm$ updated}
    \Until {$\Li = \emptyset$}
    \State \Return $\alpha$; \Cmmnt{$\scp(\lt_{i_n} | \lt_{i_m}) = \scp(\lt_{i_n} | \lt_j, \lt_{i_1\hskip-0.09em}, \ldots, \lt_{i_m})$ (see also Appendix \ref{s:App})}
\end{algorithmic}
\label{alg:CnstAssg}
\end{algorithm}

\begin{definition}\label{d:indpn}
Let $\Fm = {^{_1\hskip-0.09em}\Fm} \wedge {^{_2\hskip-0.09em}\Fm} \wedge \cdots \wedge {^{_l\hskip-0.09em}\Fm}$ such that ${^{_1\hskip-0.09em}\Fm}, {^{_2\hskip-0.09em}\Fm}, \ldots, {^{_l\hskip-0.09em}\Fm}$ are disjoint, or independent formulas. That is, ${^{_1\hskip-0.09em}\Li} \cap {^{_2\hskip-0.09em}\Li} \cap \cdots \cap {^{_l\hskip-0.09em}\Li} = \emptyset$.
\end{definition}

\begin{example}\label{e:example3}
Let ${^{_1\hskip-0.09em}\Fm} = (x_{1\hskip-0.12em} \X \nx2 \X x_6) \wedge (x_3 \X x_4 \X \nx5) \wedge (x_3 \X x_6 \X \nx7) \wedge (x_4 \X x_6 \X \nx7)$, $^{_2\hskip-0.09em}\Fm = (x_8 \X x_9 \X \nx{10})$, and $^{_3\hskip-0.09em}\Fm = (x_{11\hskip-0.09em} \X \nx{12} \X x_{13})$ to form ${\color{purple}\FM} = {^{_1\hskip-0.09em}\Fm} \wedge {^{_2\hskip-0.09em}\Fm} \wedge {^{_3\hskip-0.09em}\Fm}$ by Definition \ref{d:indpn}. Then, \texttt{Scan}~$\!(\FM_4)$ terminates, that is, $\FM$ is satisfiable. Thus, $\hat{\FM} = \hat{\scp} \wedge \hat{\Fm}$, where $\hat{\scp} = \nx3 \wedge \nx4 \wedge \nx5$ and $\hat{\Fm} = (x_{1\hskip-0.12em} \X \nx2 \X x_6) \wedge (x_6 \X \nx7) \wedge {^{_2\hskip-0.09em}\Fm} \wedge {^{_3\hskip-0.09em}\Fm}$ (see Example \ref{e:example2}). Let $\scp \coloneqq \hat{\scp}$ and ${\color{purple}\Fm} \coloneqq \hat{\Fm}$, i.e., $\Li \coloneqq \Li^{_{\hat{\Fm}\!}}$. Hence, $\Li^{_\scp\!\hskip-0.07em} = \{3, 4, 5\}$, and $\Li = \{1, 2, \ldots, 13\} - \Li^{_\scp\!\hskip-0.07em}$. Then, a satisfying assignment $\alpha$ is determined by composing $\scp(\lt_i | \lt_j)$ constructed over $\Fm'(\lt_j)$. The following shows some of the scopes $\scp(\lt_i)$ and beyond the scopes $\Fm'(\lt_i)$, constructed over $\Fm$ when the scan terminates.

\begin{tabular}{ r l l l r l l}
  $\scp(x_1)\hskip-.9em$                &$=$& $\hskip-.9em x_{1\hskip-0.12em} \wedge x_2 \wedge \nx6 \wedge \nx7$                              & $\hskip-.5em\&\hskip-.5em$ &
  $\Fm'(x_1)\hskip-.9em$                &$=$& $\hskip-.9em {^{_2\hskip-0.09em}\Fm} \wedge {^{_3\hskip-0.09em}\Fm}$                                                            \\

  $\scp(x_2)\hskip-.9em$                &$=$& $\hskip-.9em x_2$                                                                                & $\hskip-.5em\&\hskip-.5em$ &
  $\Fm'(x_2)\hskip-.9em$                &$=$& $\hskip-.9em (x_{1\hskip-0.12em} \X x_6)\wedge (x_6 \X \nx7)\wedge {^{_2\hskip-0.09em}\Fm} \wedge {^{_3\hskip-0.09em}\Fm}$  \\

  $\scp(\nx2)\hskip-.9em$               &$=$& $\hskip-.9em \nx{1\hskip-0.12em} \wedge \nx2 \wedge \nx6 \wedge \nx7$                            & $\hskip-.5em\&\hskip-.5em$ &
  $\Fm'(\nx2)\hskip-.9em$               &$=$& $\hskip-.9em {^{_2\hskip-0.09em}\Fm} \wedge {^{_3\hskip-0.09em}\Fm}$                                                            \\

  $\scp(x_6) = \scp(x_7)\hskip-.9em$                                  &$=$& $\hskip-.9em \nx{1\hskip-0.12em} \wedge x_2 \wedge x_6 \wedge x_7$ & $\hskip-.5em\&\hskip-.5em$ & $\Fm'(x_6) = \Fm'(x_7)\hskip-.9em$    &$=$& $\hskip-.9em{^{_2\hskip-0.09em}\Fm} \wedge {^{_3\hskip-0.09em}\Fm}$                                                             \\

  $\scp(\nx6) = \scp(\nx7)\hskip-.9em$  &$=$& $\hskip-.9em \nx6 \wedge \nx7$                                                                   & $\hskip-.5em\&\hskip-.5em$ &
  $\Fm'(\nx6) = \Fm'(\nx7)\hskip-.9em$  &$=$& $\hskip-.9em (x_{1\hskip-0.12em} \X \nx2) \wedge {^{_2\hskip-0.09em}\Fm} \wedge {^{_3\hskip-0.09em}\Fm}$                      \\

  $\scp(x_8)\hskip-.9em$                &$=$& $\hskip-.9em x_8 \wedge \nx9 \wedge x_{10}$                                                      & $\hskip-.5em\&\hskip-.5em$ &
  $\Fm'(x_8)\hskip-.9em$                &$=$& $\hskip-.9em (x_{1\hskip-0.12em} \X \nx2 \X x_6) \wedge (x_6 \X \nx7) \wedge {^{_3\hskip-0.09em}\Fm}$                         \\

  $\scp(x_{11})\hskip-.9em$             &$=$& $\hskip-.9em x_{11\hskip-0.09em} \wedge x_{12} \wedge \nx{13}$                                   & $\hskip-.5em\&\hskip-.5em$ &
  $\Fm'(x_{11})\hskip-.9em$             &$=$& $\hskip-.9em (x_{1\hskip-0.12em} \X \nx2 \X x_6) \wedge (x_6 \X \nx7) \wedge {^{_2\hskip-0.09em}\Fm}$
\end{tabular}
\end{example}

\begin{example}
A satisfying assignment $\alpha$ is constructed by an order of indices over $\Li$, $\Li = \{1, \ldots, 13\} - \Li^{_\scp\!\hskip-0.07em}$ (Example \ref{e:example3}), such that $\lt_{i\hskip-0.09em} \coloneqq x_{i\hskip-0.09em}$ for any $\scp(\lt_i)$ throughout the construction. First, pick $6 \in \Li$. As a result, $\alpha \gets \scp(x_6)$ and $\Li \gets \Li - \Li(x_6)$, where $\scp(x_6) = \{\nx{1\hskip-0.07em}, x_2, x_6, x_7\}$, $\Li(x_6) = \{1, 2, 6, 7\}$, and $\Li \gets \{8, 9, 10, 11, 12, 13\}$. Then, pick 8, hence $\alpha \gets \alpha \cup \scp(x_8 | x_6)$, where $\scp(x_8 | x_6) = \{x_8, \nx9, x_{10}\}$. Also, $\Li \gets \Li - \Li(x_8 | x_6)$, where $\Li(x_8 | x_6) = \{8, 9, 10\}$, hence $\Li \gets \{11, 12, 13\}$. Finally, pick 11. Therefore, $\alpha \gets \alpha \cup \scp(x_{11\hskip-0.05em} | x_6, x_8)$ such that $\Li \gets \emptyset$, which indicates its termination. Note that \texttt{Scope}~$\!\big(x_{11\hskip-0.07em}, \Fm'(x_8 | x_6)\big)$ constructs $\scp(x_{11\hskip-0.05em} | x_6, x_8)$, in which $\Fm'(x_8 | x_6) = {^{_3\hskip-0.09em}\Fm}$, and that $\Li'(x_{11\hskip-0.05em} | x_6, x_8) = \emptyset$ iff $\Li \gets \emptyset$. Note also that $\scp(x_8 | x_6) = \scp(x_8)$ and $\scp(x_{11\hskip-0.05em} | x_6, x_8) = \scp(x_{11})$, since $^{_1\hskip-0.07em}\Fm$, ${^{_2\hskip-0.05em}\Fm}$ and ${^{_3\hskip-0.05em}\Fm}$ are disjoint by Definition \ref{d:indpn}. Consequently, Algorithm \ref{alg:CnstAssg} constructs $\alpha = \{\scp(x_6), \scp(x_8 | x_6), \scp(x_{11\hskip-0.05em} | x_6, x_8)\}$. Note that $\FM$ is \emph{decomposed} into $\scp$, $\scp(x_6)$, $\scp(x_8 | x_6)$, and $\scp(x_{11\hskip-0.05em} | x_6, x_8)$, which are \emph{disjoint} (see also Note \ref{n:Trmnts} and Lemma \ref{l:prtn}).
\end{example}

\begin{example}
Let $(2, 1, 8, 11)$ be another order of indices in Example \ref{e:example3}. This order leads to the assignment $\{\scp, \scp(x_2), \scp(x_1 | x_2), \scp(x_8 | x_2, x_1), \scp(x_{11\hskip-0.05em} | x_2, x_{1\hskip-0.07em}, x_8)\}$ for $\FM$. This assignment corresponds to the partition $\big\{\Li^{_\scp\!}, \{2\}, \{1,6,7\}, \{8,9,10\}, \{11,12,13\}\big\}$, where $\Li^{_\scp\!\hskip-0.07em} = \{3, 4, 5\}$ (see also Note \ref{n:Lprt} and Lemma \ref{l:prtn}). Note that the scope $\scp(x_1)$ is constructed over $\Fm$, and the conditional scope $\scp(x_1 | x_2)$ is constructed over $\Fm'(x_2)$, where $\Fm \supseteq \Fm'(x_2)$. Recall that $\Fm \coloneqq \hat{\Fm}$. Hence, $\scp(x_1) \ent \scp(x_1 | x_2)$, in which $\scp(x_1) = x_{1\hskip-0.12em} \wedge x_2 \wedge \nx6 \wedge \nx7$, while $\scp(x_1 | x_2) = x_{1\hskip-0.12em} \wedge \nx6 \wedge \nx7$. Moreover, $\scp(x_8) \ent \scp(x_8 | x_2, x_1)$ due to $\Fm \supseteq \Fm'(x_1 | x_2)$, and $\scp(x_{11}) \ent \scp(x_{11\hskip-0.03em} | x_2, x_{1\hskip-0.07em}, x_8)$ due to $\Fm \supseteq \Fm'(x_8 | x_2, x_1)$, where $\Fm'(x_1 | x_2) = {^{_2\hskip-0.09em}\Fm} \wedge {^{_3\hskip-0.09em}\Fm}$ and $\Fm'(x_8 | x_2, x_1) = {^{_3\hskip-0.09em}\Fm}$ (see Lemmas \ref{l:SubSet}-\ref{l:EntGn}).
\end{example}

\subsection[Example]{An Illustrative Example}
This section illustrates \texttt{Scan}~$\!(\FM_s)$. Let $\FM \hskip-0.03em= \Fm \hskip-0.03em= (x_{1\hskip-0.12em} \X \nx3) \wedge (x_{1\hskip-0.12em} \X \nx2 \X x_3) \wedge (x_2 \X \nx3)$, which is adapted from Esparza~\cite{Esp98}, and denotes a general formula by Definition \ref{d:GnFm}. Note that $C_{\color{green!40!black}1\hskip-0.12em} = \linebreak \{x_{\color{red}1\hskip-0.07em}, \nx{\color{red}3}\}$, $C_{\color{green!40!black}2\hskip-0.09em} = \{x_{\color{red}1\hskip-0.07em}, \nx{\color{red}2}, x_{\color{red}3}\}$, and $C_{\color{green!40!black}3\hskip-0.09em} = \{x_{\color{red}2}, \nx{\color{red}3}\}$. Hence, $\Ckk =\{{\color{green!40!black}1, 2, 3}\}$, and $\Li = \Li^{_\Fm\!} =\{{\color{red}1, 2, 3}\}$.

\texttt{Scan}~$\!(\FM)$: There exists no conjunct in (the initial formula) $\FM$. That is, $\scp$ is empty (\hyperref[alg:Scan]{L:1}). Recall that $\FM \coloneqq \FM_{1\hskip-0.07em}$, and that $\lt_{i\hskip-0.09em} \in \{x_i, \nx i\}$. Recall also that \emph{nontrivial} incompatibility of $\lt_{i\hskip-0.09em}$ is checked (\hyperref[alg:Scan]{L:4-8}) via \texttt{Scope}~$\!(\lt_i, \Fm)$. Moreover, the order of incompatibility check is arbitrary (incompatibility is monotonic) by Theorem \ref{t:mntn}. Let \texttt{Scope}~$\!(x_{1\hskip-0.07em}, \Fm)$ execute due to \texttt{Scan} \hyperref[alg:Scan]{L:6}.

\texttt{Scope}~$\!(x_{1\hskip-0.07em}, \Fm)$: Since $\scp(x_1) \supseteq \{x_3, \nx3\}$, $x_{1\hskip-0.12em}$ is incompatible \emph{nontrivially} (see Example \ref{e:Scope}). Thus, $\nx{1\hskip-0.17em}$ becomes necessary (a conjunct). Then, \texttt{Remove}~$\!(x_{1\hskip-0.07em}, \Fm)$ executes due to \texttt{Scan} \hyperref[alg:Scan]{L:6}.

\texttt{Remove}~$\!(x_{\color{red}1\hskip-0.07em}, \Fm)$: $\Ckk^{\nx1\hskip-0.07em} \!= \emptyset$ by \texttt{OvrlEft} \hyperref[alg:Ovrl]{L:1}. $\Ckk^{x_1\hskip-0.07em} \!= \{1, 2\}$, thus $\Fm^{x_1\hskip-0.07em} \!= (x_{1\hskip-0.12em} \X \nx3) \wedge (x_{1\hskip-0.12em} \X \nx2 \X x_3)$ by \texttt{OvrlEft} \hyperref[alg:Ovrl]{L:7}. As a result, $\lcl{}(\nx1) = \{\nx3\}$ \& $\lcln(\neg x_1) = \big\{\{\}, \{\nx2, x_3\}\big\}$, the effects of $\nx{1\hskip-0.12em}$ and $\neg x_{1\hskip-0.07em}$. Note that $C_{1\!} \gets \emptyset$. Then, $\scp_{2\hskip-0.07em} \gets \scp \cup \{\nx1\} \cup \lcl{}(\nx1)$ (\texttt{Remove} \hyperref[alg:Remove]{L:2}), and $\Li^{_\Fm\!} \gets \Li^{_\Fm\!} - \{{\color{red}1}\}$ and $\Li^{_\scp\!\hskip-0.07em} \gets \Li^{_\scp\!} \cup \{{\color{red}1}\}$ (\hyperref[alg:Remove]{L:4}). Also, $\Fm_{2\hskip-0.07em} \gets \lcln(\neg x_1) \wedge \Fm'$, where $\lcln(\neg x_1) = (\nx{2\hskip-0.05em} \X x_3)$ and $\Fm' \hskip-0.12em= (x_{2\hskip-0.05em} \X \nx3)$ (\hyperref[alg:Remove]{L:5}). As a result, $\scp_{2\hskip-0.07em} = \nx{1\hskip-0.11em} \wedge \nx3$, and $\Fm_{2\hskip-0.07em} = (\nx{2\hskip-0.05em} \X x_3) \wedge (x_{2\hskip-0.05em} \X \nx3)$. Note that $C_{\color{blue}1\hskip-0.12em} = \{\nx2, x_3\}$ and $C_{\color{blue}2} = \{x_2, \nx3\}$. Consequently, $\FM_{2\hskip-0.05em} = \scp_{2\hskip-0.05em} \wedge \Fm_{2\hskip-0.03em}$, and  \texttt{Scan}~$\!(\FM_2)$ executes due to \texttt{Remove} \hyperref[alg:Remove]{L:6}.

\texttt{Scan}~$\!(\FM_2)$: $\Ckk_{2\hskip-0.05em} = \{{\color{blue}1, 2}\}$ and $\Li^{_\Fm\!} = \{2, 3\}$ hold in $\Fm_2$. Then, $\{x_2, \nx2\} \cap \scp_2 = \emptyset$ for $2 \in \Li^{_\Fm\!}$, while $\nx3 \hskip-0.05em\in \scp_2$ for $3 \in \Li^{_\Fm\!}$ (\hyperref[alg:Scan]{L:1}). As a result, $\nx{3\hskip-0.05em}$ is \emph{necessary} for satisfying $\FM_2$, hence $\nx{3\hskip-0.09em} \imp \neg x_3$, that is, $x_{3\hskip-0.05em}$ is incompatible \emph{trivially}. Then, \texttt{Remove}~$\!(x_3, \Fm_2)$ executes due to \texttt{Scan} \hyperref[alg:Scan]{L:2}.

\texttt{Remove}~$\!(x_{\color{red}3}, \Fm_2)$: $\Ckk_2^{\nx3} \!= \{2\}$, thus $\Fm_2^{\nx3} \!= ({\color{white!30!blue}x_2} \hskip-0.07em\X \nx3)$, and $\Ckk_2^{x_3} \!= \{1\}$, thus $\Fm_2^{x_3} \!=  ({\color{brown}\nx2} \hskip-0.07em\X x_3)$. As a result, $\lcl{2}(\nx3) = \{{\color{white!30!blue}\nx2}\} \cup \{{\color{brown}\nx2}\}$ \& $\lcln_2(\neg x_3) = \big\{\{\}\big\}$, because $C_{1\!} = \{{\color{brown}\nx2}\}$ consists in $\lcl{2}(\nx3)$, rather than in $\lcln_2(\neg x_3)$ (see \texttt{OvrlEft} \hyperref[alg:Ovrl]{L:9}). Hence, $\scp_{3\hskip-0.07em} \gets \scp_{2\hskip-0.03em} \cup \{\nx3\} \cup \lcl{2}(\nx3)$, $\Li^{_\Fm\!} \gets \Li^{_\Fm\!} - \{{\color{red}3}\}$, and $\Li^{_\scp\!\hskip-0.07em} \gets \Li^{_\scp\!} \cup \{{\color{red}3}\}$, i.e., $\Li^{_\Fm\!} = \{2\}$. Therefore, $\Fm_{3\hskip-0.07em} = \big\{\{\}\big\}$, thus $\Ckk_{3\hskip-0.09em} = \emptyset$, and $\scp_{3\hskip-0.07em} = \nx{1\hskip-0.12em} \wedge \nx3 \wedge \nx2$.

\texttt{Scan}~$\!(\FM_3)$: $\nx2 \hskip-0.05em\in \scp_3$ for $2 \in \Li^{_\Fm\!}$ over $\Fm_3$. Then, \texttt{Remove}~$\!(x_2, \Fm_3)$ executes due to \texttt{Scan} \hyperref[alg:Scan]{L:2}.

\texttt{Remove}~$\!(x_2, \Fm_3)$: $\lcl{3}(\nx2) = \emptyset$ \& $\lcln_3(\neg x_2) = \big\{\{\}\big\}$ due to \texttt{OvrlEft}~$\!(\nx2, \Fm_3)$, because $\Ckk_3^{\nx2} \!= \emptyset$ and $\Ckk_3^{x_2} \!= \emptyset$, since $\Ckk_{3\hskip-0.09em} = \emptyset$. Hence, $\Li^{_\Fm\!} \gets \{2\} - \{2\}$ and $\Fm_4 \gets \Fm_3$. Then, \texttt{Scan}~$\!(\FM_4)$ executes.

\texttt{Scan}~$\!(\FM_4)$ \emph{terminates}: $\hat{\FM} \hskip-0.05em= \hat{\scp} \hskip-0.05em= \nx{1\hskip-0.13em} \wedge \nx{3\hskip-0.07em} \wedge \nx2$ (\hyperref[alg:Scan]{L:9}), and $\FM$ collapses to a unique assignment.

Let \texttt{Scope}~$\!(x_3, \Fm)$ execute \emph{before} \texttt{Scope}~$\!(x_{1\hskip-0.07em}, \Fm$) due to \texttt{Scan} \hyperref[alg:Scan]{L:6} (see Theorem \ref{t:mntn}).

\texttt{Scope}~$\!(x_3, \Fm)$: $\scp(x_3) \gets \{x_3\}$ and $\Fm_* \gets \Fm$ (\hyperref[alg:Scope]{L:1}). Then, $\Ckk_*^{x_3} \!=\{{\color{red}2}\}$ due to \texttt{OvrlEft}~$\!(x_3, \Fm_*)$ \hyperref[alg:Ovrl]{L:1}, hence $\Fm_*^{x_3} \!= (x_{1\hskip-0.12em} \X \nx2 \X x_3)$. As a result, $c_{\color{red}2} \gets \{\nx{1\hskip-0.07em}, x_2\}$ and $\lcl{*}(x_3) \gets \lcl{*}(x_3) \cup c_{\color{red}2}$ (\hyperref[alg:Ovrl]{L:3,5}). Moreover, $\Ckk_*^{\nx3} \!= \{1, 3\}$ (\hyperref[alg:Ovrl]{L:7}), hence $\Fm_*^{\nx3\hskip-0.09em} = (x_{1\hskip-0.12em} \X \nx3) \wedge (x_{2\hskip-0.05em} \X \nx3)$. Then, $C_{1\hskip-0.12em} \gets \{x_{1\hskip-0.07em}, \nx3\} - \{\nx3\}$, $\lcl{*}(x_3) \gets \lcl{*}(x_3) \cup C_{1\hskip-0.09em}$, and $C_{1\hskip-0.12em} \gets \emptyset$. Likewise, $C_{3\hskip-0.07em} \gets \{x_2, \nx3\} - \{\nx3\}$, $\lcl{*}(x_3) \gets \lcl{*}(x_3) \cup C_{3\hskip-0.05em}$, and $C_{3\hskip-0.09em} \gets \emptyset$ (\texttt{OvrlEft} \hyperref[alg:Ovrl]{L:8-9}). Consequently, $\lcl{*}(x_3) \gets \{\nx{1\hskip-0.07em}, x_2, x_1\}$ \& $\lcln_*(\neg \nx3) \gets \Fm_*^{\nx3}$  (\hyperref[alg:Ovrl]{L:11}). Note that $\Fm_*^{\nx3} \!= \big\{\{\},\{\} \big\}$, since $C_{1\hskip-0.12em} =  C_{3\hskip-0.09em} = \emptyset$. Then, $\scp(x_3) \gets \scp(x_3) \cup \{x_3\} \cup \lcl{*}(x_3)$ due to \texttt{Scope} \hyperref[alg:Scope]{L:4}, hence $\scp(x_3) = \{x_3, \nx{1\hskip-0.07em}, x_2, x_1\}$. Since $\scp(x_3) \supseteq \{\nx{1\hskip-0.07em}, x_1\}$ (\hyperref[alg:Scope]{L:5}), $x_3$ is incompatible \emph{nontrivially}, i.e., $x_{3\hskip-0.09em} \imp \nx{1\hskip-0.12em} \wedge x_{1\hskip-0.12em}$ and $\neg x_{3\hskip-0.09em} \imp \nx3$. Then, \texttt{Remove}~$\!(x_3, \Fm)$ executes due to \texttt{Scan} \hyperref[alg:Scan]{L:6}.

\texttt{Remove}~$\!(x_{\color{red}3}, \Fm)$: $\Fm^{\nx3} \!= (x_{1\hskip-0.12em} \X \nx3) \wedge (x_{2\hskip-0.05em} \X \nx3)$ due to $\Ckk^{\nx3} \!= \{1, 3\}$, and $\Fm^{x_3} \!= (x_{1\hskip-0.12em} \X \nx2 \X x_3)$ due to $\Ckk^{x_3} \!= \{2\}$. Then, \texttt{OvrlEft}~$\!(\nx3, \Fm)$ returns $\lcl{}(\nx3) = \{\nx{1\hskip-0.07em}, \nx2\}$ \& $\lcln(\neg x_3) = \big\{\{x_{1\hskip-0.07em}, \nx2\}\big\}$ (\texttt{Remove} \hyperref[alg:Remove]{L:1}), $\scp_{2\hskip-0.09em} \gets \scp \cup \{\nx3\} \cup \lcl{}(\nx3)$ (\hyperref[alg:Remove]{L:2}), and $\Li^{_\Fm\!} \gets \Li^{_\Fm\!} - \{{\color{red}3}\}$ and $\Li^{_\scp\!\hskip-0.07em} \gets \Li^{_\scp\!} \cup \{{\color{red}3}\}$ (\hyperref[alg:Remove]{L:4}). As a result, $\scp_{2\hskip-0.09em} = \nx{3\hskip-0.07em} \wedge \nx{1\hskip-0.12em} \wedge \nx{2\hskip-0.03em}$. Moreover, $\Fm_{2\hskip-0.07em} \gets \lcln(\neg x_3) \wedge \Fm'\hskip-0.15em$ (\hyperref[alg:Remove]{L:5}), in which $\lcln(\neg x_3) = (x_{1\hskip-0.12em} \X \nx2)$ and $\Fm'\hskip-0.15em$ is empty. Therefore, $\FM_{2\hskip-0.07em} = \scp_{2\hskip-0.07em} \wedge \Fm_{2\hskip-0.07em}$. Note that $C_{\color{orange}1\hskip-0.12em} = \{x_{1\hskip-0.07em}, \nx2\}$, hence $\Ckk_{2\hskip-0.05em} = \{{\color{orange}1}\}$. Recall that $\Li^{_\Fm\!} = \{1, 2\}$, and that $\Li^{_\scp\!\hskip-0.07em} = \{{\color{red}3}\}$. Then, \texttt{Scan}~$\!(\FM_2)$ executes due to \texttt{Remove}~$\!(x_3, \Fm)$ \hyperref[alg:Remove]{L:6}.

\texttt{Scan}~$\!(\FM_2)$: $\Li^{_\Fm\!} = \{1, 2\}$ such that $\nx2 \in \scp_2$ and $\nx{1\!} \in \scp_2$. Thus, $\nx{2\hskip-0.05em}$ and $\nx{1\hskip-0.12em}$ are \emph{necessary}, hence $x_{2\hskip-0.05em}$ and $x_{1\hskip-0.12em}$ are incompatible \emph{trivially}. Then, \texttt{Remove}~$\!(x_{1\hskip-0.08em}, \Fm_2)$ and \texttt{Remove}~$\!(x_2, \Fm_2)$ execute.

The fact that the order of incompatibility check is arbitrary (Theorem \ref{t:mntn}) is illustrated as follows. \texttt{Scope}~$\!(x_3, \Fm)$ returns $x_{3\hskip-0.09em}$ is incompatible \emph{nontrivially}, since $x_{3\hskip-0.09em} \imp \nx{1\hskip-0.15em} \wedge x_{1\hskip-0.07em}$. Therefore, $\neg \nx{1\hskip-0.19em} \vee \neg x_{1\hskip-0.12em} \imp \neg x_{3\hskip-0.03em}$, hence $x_{1\hskip-0.19em} \vee \nx{1\hskip-0.12em} \imp \nx{3\hskip-0.03em}$. Then, $\nx{3\hskip-0.09em} \imp \nx{1\hskip-0.12em}$ due to $C_{1\hskip-0.12em} = (x_{1\hskip-0.12em} \X \nx3)$, and $\nx{1\hskip-0.09em} \imp \neg x_{1\hskip-0.12em}$. Thus, $x_{1\hskip-0.12em}$ is \emph{still} incompatible, but trivially \big(cf. \texttt{Scope}~$\!(x_{1\hskip-0.07em}, \Fm)$\big), even if $\neg x_{3\hskip-0.07em}$ holds. That is, $x_{1\hskip-0.12em}$ the \emph{nontrivial} incompatible in $\Fm$ due to $x_{1\hskip-0.15em} \imp \nx{3\hskip-0.09em} \wedge x_{3\hskip-0.05em}$, i.e., $\neg \nx{3\hskip-0.12em} \vee \neg x_{3\hskip-0.10em} \imp \neg x_{1\hskip-0.07em}$, is incompatible \emph{trivially} in $\scp_2$ due to $\nx{1\hskip-0.09em} \imp \neg x_{1\hskip-0.12em}$. See \texttt{Scan}~$\!(\FM_2)$ above. Also, since $x_{3\hskip-0.07em} \notin C_{k\hskip-0.09em}$ and $\nx{3\hskip-0.07em} \notin C_{k\hskip-0.09em}$ in $\Fm_{s\hskip-0.07em}$ for any $s \geqslant 2$, $\unsat \FM_s(x_3)$ for all $s \geqslant 2$, even if any $\lt_{i\hskip-0.12em}$ is removed from some $C_{k\hskip-0.07em}$ in $\Fm_{s\hskip-0.05em}$, $s \geqslant 2$.

\section{Conclusion}

X3SAT has proved to be effective to show $\PP = \NP$. A polynomial time algorithm checks unsatisfiability of $\Fm(\lt_i)$ such that $\unsat \Fm(\lt_i)$ iff $\scp_s(\lt_i)$ involves $x_{j\hskip-0.09em} \wedge \nx{j\hskip-0.09em}$ for some $s$. Thus, $\Fm(\lt_i)$ reduces to $\scp(\lt_i)$. $\scp(\lt_i)$ denotes a conjunction of literals that are \emph{true}, since each $\lt_{j\hskip-0.12em}$ such that $\unsat \scp_s(\lt_j)$ is removed from $\Fm$. Hence, $\Fm$ is satisfiable iff $\scp(\lt_i)$ is satisfied for any $\lt_{i\hskip-0.09em} \in \{x_i, \nx{i}\}$. Thus, it is \emph{easy} to verify satisfiability of $\Fm$ via satisfiability of $\scp(x_1), \scp(\nx1), \ldots, \scp(x_n), \scp(\nx n)$.

\bibliography{PvsNP}

\appendix
\section{Proof of Theorem \ref{t:InCmp}/\ref{c:Sat}}\label{s:App}

This section gives a rigorous proof of Theorem \ref{t:InCmp}/\ref{c:Sat}. Recall that the $\FM_{s\hskip-0.05em}$ scan is \emph{interrupted} iff $\scp_{s\hskip-0.07em}$ involves $x_{i\hskip-0.09em} \wedge \nx{i\hskip-0.09em}$ for some $i$ and $s$, that is, $\FM$ is unsatisfiable, which is trivial to verify. Recall also that the $\FM_{\hat{s}\hskip-0.05em}$ scan \emph{terminates} iff $\scp_{\hat{s}}(\lt_i) = \mathbf{T}$ for any $i \hskip-0.05em \in \Li^{_{\hat{\Fm}\!}}$, $\lt_{i\hskip-0.09em} \in \{x_i, \nx{i}\}$. Moreover, $\hat{\FM} \hskip-0.05em= \hat{\scp} \wedge \hat{\Fm}$ such that $\hat{\scp} \hskip-0.05em= \mathbf{T}$ (see \texttt{Scan} \hyperref[alg:Scan]{L:9} and Note \ref{n:Trmnts}). Therefore, when the scan terminates, satisfiability of $\hat{\Fm}$ is to be proved, which is addressed in this section. Let $\Fm \hskip-0.05em\coloneqq \hat{\Fm}$, i.e., $\Li \hskip-0.05em\coloneqq \Li^{_{\hat{\Fm}\!}}$.

\begin{theorem}[cf. \!\ref{t:InCmp}-\ref{c:Sat}/Claim \ref{cl:first}]\label{t:final}
These statements are equivalent for any $i \in \Li$: $a) \unsat \Fm(\lt_i)$ iff $\unsat \scp_s(\lt_i)$ for some $s$. $b) \,\lt_{i\hskip-0.12em} \ent \scp(\lt_i)$. $c) \,\sat \Fm$ by $\alpha = \{\scp(\lt_{i_0}), \scp(\lt_{i_1\hskip-0.05em} | \lt_{i_0}), \ldots, \scp(\lt_{i_n} | \lt_{i_m})\}$.
\end{theorem}
\begin{proof}
We will show $a \imp b$, $b \imp c$, and $c \imp a$ (see Kenneth H. Rosen, Discrete Mathematics and its Applications, 7E, pg. 88). Firstly, $a \imp b$ holds, because $a$ holds by assumption (see Note \ref{n:assmp}), and $b$ holds by Lemma \ref{l:Scope}. Next, we will show $b \imp c$. We do this by showing that satisfiability of $\Fm$ is \emph{preserved} throughout the assignment $\alpha$ construction, where $\alpha = \{\scp(\lt_{i_0}), \scp(\lt_{i_1\hskip-0.05em} | \lt_{i_0}), \ldots, \scp(\lt_{i_n} | \lt_{i_m})\}$, because any \emph{partial} assignment $\scp(\lt_i | \lt_j)$ is constructed \emph{arbitrarily} through consecutive steps having the Markov property. Thus, construction of $\scp(\lt_i | \lt_j)$ in the next step is independent from the preceding steps, and depends only upon $\scp(\lt_j | \lt_k)$ in the present step (see also Lemma \ref{l:prtn}). The construction process is specified below.

\textsf{Step} 0: Pick any $\lt_{i_0\hskip-0.20em}$ in $\Fm$. Then, $\lt_{i_0\hskip-0.12em} \ent \scp(\lt_{i_0})$ by Lemma \ref{l:Scope}. Also, $\lt_{i_0\hskip-0.15em}$ partitions \color{purple}$\Li$ into $\Li(\lt_{i_0})$ and $\Li'(\lt_{i_0})$. \color{black} Note that $i_0\hskip-0.03em \in \Li$ and $i_0\hskip-0.03em \in \Li(\lt_{i_0})$. Hence, $i_0\hskip-0.03em \notin \Li'(\lt_{i_0})$ by Lemma \ref{l:Indp}. Therefore, $\Fm(\lt_{i_0}) = \scp(\lt_{i_0}) \wedge \Fm'(\lt_{i_0})$ in \textsf{Step} 0. Then, pick an \emph{arbitrary} $\lt_{i_1\hskip-0.20em}$ in $\Fm'(\lt_{i_0})$ for \textsf{Step} 1.

\textsf{Step} 1: $\color{purple}\Li(\lt_{i_0}) \cap \Li'(\lt_{i_0}) = \emptyset$ due to \textsf{Step} 0. Then, $\lt_{i_1\hskip-0.12em} \ent \scp(\lt_{i_1})$ by Lemma \ref{l:Scope}, as well as  $\scp(\lt_{i_1}) \ent \scp(\lt_{i_1} | \lt_{i_0})$ by Lemma \ref{l:Ent}. Also, $\lt_{i_1\hskip-0.20em}$ partitions $\color{purple}\Li'(\lt_{i_0})$ into $\Li(\lt_{i_1} | \lt_{i_0})$ and $\Li'(\lt_{i_1} | \lt_{i_0})$. Thus, $\Li(\lt_{i_0}) \cap \Li(\lt_{i_1\hskip-0.05em} | \lt_{i_0}) = \emptyset$, since $\Li'(\lt_{i_0}) \supseteq \Li(\lt_{i_1\hskip-0.05em} | \lt_{i_0})$. As a result, $\Li$ is partitioned into $\Li(\lt_{i_0})$, $\Li(\lt_{i_1\hskip-0.07em} | \lt_{i_0})$, and $\Li'(\lt_{i_1\hskip-0.07em} | \lt_{i_0})$ by $\lt_{i_0\hskip-0.10em}$ and $\lt_{i_1\hskip-0.15em}$. Thus, $\scp(\lt_{i_0})$ and $\scp(\lt_{i_1\hskip-0.05em} | \lt_{i_0})$ are \emph{disjoint}, as well as \emph{true}. Therefore, $\color{white!30!blue}\scp(\lt_{i_0}) \wedge \scp(\lt_{i_1\hskip-0.05em} | \lt_{i_0}) = \mathbf{T},$ and $\Fm(\lt_{i_0}, \lt_{i_1}) = \scp(\lt_{i_0}) \wedge \scp(\lt_{i_1\hskip-0.05em} | \lt_{i_0}) \wedge \Fm'(\lt_{i_1\hskip-0.05em} | \lt_{i_0})$.

\textsf{Step} 2: The preceding steps have partitioned $\Li$ into $\Li(\lt_{i_0}) \cup \Li(\lt_{i_1\hskip-0.05em} | \lt_{i_0})$ and $\Li'(\lt_{i_1\hskip-0.05em} | \lt_{i_0})$.  Then, $\lt_{i_2\hskip-0.12em} \ent \scp(\lt_{i_2})$ by Lemma \ref{l:Scope}, as well as  $\scp(\lt_{i_2}) \ent \scp(\lt_{i_2} | \lt_{i_1})$ by Lemma \ref{l:Ent}/\ref{l:EntGn}. Also, $\lt_{i_2\hskip-0.12em}$ in $\Fm'(\lt_{i_1\hskip-0.05em} | \lt_{i_0})$ partitions $\Li'(\lt_{i_1\hskip-0.05em} | \lt_{i_0})$ into $\Li(\lt_{i_2\hskip-0.03em} | \lt_{i_1\hskip-0.05em})$ and $\Li'(\lt_{i_2\hskip-0.03em} | \lt_{i_1\hskip-0.05em})$, i.e., $\Li'(\lt_{i_1\hskip-0.05em} | \lt_{i_0}) \supseteq \Li(\lt_{i_2\hskip-0.03em} | \lt_{i_1\hskip-0.05em})$. Then, $\big(\Li(\lt_{i_0}) \cup \Li(\lt_{i_1\hskip-0.05em} | \lt_{i_0})\big) \cap \Li(\lt_{i_2} | \lt_{i_1\hskip-0.05em}) = \emptyset$, thus $\color{white!30!blue}\scp(\lt_{i_0}) \wedge \scp(\lt_{i_1\hskip-0.05em} | \lt_{i_0})$ and $\scp(\lt_{i_2} | \lt_{i_1\hskip-0.05em})$ are \emph{disjoint}, as well as \emph{true}. Therefore, $\Fm(\lt_{i_0}, \lt_{i_1\hskip-0.07em}, \lt_{i_2}) = \scp(\lt_{i_0}) \wedge \scp(\lt_{i_1\hskip-0.05em} | \lt_{i_0}) \wedge \scp(\lt_{i_2} | \lt_{i_1\hskip-0.05em}) \wedge \Fm'(\lt_{i_2} | \lt_{i_1\hskip-0.05em})$, in which $\scp(\lt_{i_0}) \wedge \scp(\lt_{i_1\hskip-0.05em} | \lt_{i_0}) \wedge \scp(\lt_{i_2} | \lt_{i_1\hskip-0.05em}) \hskip-0.05em=\hskip-0.05em \mathbf{T}$. Note that $\alpha \supseteq \{\scp(\lt_{i_0}), \scp(\lt_{i_1\hskip-0.05em} | \lt_{i_0}), \scp(\lt_{i_2} | \lt_{i_1\hskip-0.05em})\}$, and that $\Li$ is partitioned into $\Li(\lt_{i_0})$, $\Li(\lt_{i_1\hskip-0.07em} | \lt_{i_0})$, $\Li(\lt_{i_2\hskip-0.07em} | \lt_{i_1})$, and $\Li'(\lt_{i_2\hskip-0.07em} | \lt_{i_1})$ such that $\Li'(\lt_{i_2\hskip-0.07em} | \lt_{i_1}) \neq \emptyset$.

\textsf{Step} $n$: $\lt_{i_n\hskip-0.15em}$ partitions $\Li'(\lt_{i_m} | \lt_{i_l})$ into $\Li(\lt_{i_n} | \lt_{i_m})$ and $\Li'(\lt_{i_n} | \lt_{i_m})$ such that $\Li'(\lt_{i_n} | \lt_{i_m}) = \emptyset$. $\Li(\lt_{i_0}) \cup \Li(\lt_{i_1\hskip-0.05em} | \lt_{i_0}) \cup \cdots \cup \Li(\lt_{i_m} | \lt_{i_l})$ and $\Li'(\lt_{i_m} | \lt_{i_l})$, hence $\Li(\lt_{i_n} | \lt_{i_m})$, form a partition of $\Li$. Therefore, $\scp(\lt_{i_0}) \wedge \scp(\lt_{i_1\hskip-0.05em} | \lt_{i_0}) \wedge \cdots \wedge \scp(\lt_{i_m} | \lt_{i_l})$ and $\scp(\lt_{i_n} | \lt_{i_m})$ are \emph{disjoint}, as well as \emph{true}. That is, $\Fm(\lt_{i_0}, \lt_{i_1}, \ldots, \lt_{i_m}, \lt_{i_n}) = \scp(\lt_{i_0}) \wedge \scp(\lt_{i_1\hskip-0.05em} | \lt_{i_0}) \wedge \cdots \wedge \scp(\lt_{i_m} | \lt_{i_l}) \wedge \scp(\lt_{i_n} | \lt_{i_m})$ is satisfied.

Thus, $\Fm$ is composed of $\scp(.)$ \emph{disjoint} and \emph{satisfied}, hence $\Fm$ is satisfiable, and $b \imp c$ holds. Finally, we show $c \imp a$. $\lt_{i\hskip-0.12em}$ transforms $\Fm$ into $\scp(\lt_i) \wedge \Fm'(\lt_i)$. Then, $\Fm \equiv \scp(\lt_i) \wedge \Fm'(\lt_i)$, where $\Fm$ and $\scp(\lt_i)$ are \emph{satisfiable}, and $\scp(\lt_i)$ and $\Fm'(\lt_i)$ are \emph{disjoint}. Thus, $\Fm'(\lt_i)$ is satisfiable. Hence, unsatisfiability of $\scp_s(\lt_i)$ for some $s$ is necessary and sufficient for $\unsat \Fm_s(\lt_i)$ for any $s$.
\end{proof}

\begin{note*}
The assignment $\alpha$ construction is driven by partitioning the set $\Li'(.)$ such that $\Li \gets \Li - \Li(\lt_{i_0})$ in \textsf{Step} 1, and $\Li \gets \Li - \Li(\lt_{i_{n-1}} | \lt_{i_{n-2}})$  for $i_n\hskip-0.12em \in \Li'(\lt_{i_{n-1}} | \lt_{i_{n-2}})$ in \textsf{Step} $n \geqslant 2$.
\end{note*}

\begin{note*}
$\scp(\lt_i) \equiv \Fm(\lt_i)$ by Theorem \ref{t:final}. Thus, the formula $\Fm = \hskip-0.11em\bigwedge_{k \in \Ckk} C_{k\hskip-0.05em}$ transforms into the formula $\Fm' = \bigwedge_{i \in \Li} \mathcal{C}_{i\hskip-0.05em}$, where $C_{k\hskip-0.09em} = (\lt_{i\hskip-0.09em} \X \lt_{j\hskip-0.09em} \X \lt_v)$ and $\mathcal{C}_{i\hskip-0.11em} = \big(\scp(x_i) \oplus \scp(\nx i)\big)$. See also Note \ref{n:Trmnts}.
\end{note*}

\begin{note*}[Construction of $\alpha$]
In order to form a partition over the set $\Fm$, $\alpha$ is constructed such that $\scp(\lt_{i_1} | \lt_{i_0}) = \scp(\lt_{i_1}) - \scp(\lt_{i_0})$, and $\scp(\lt_{i_n} | \lt_{i_{n-1}}) = \scp(\lt_n) - \big(\scp(\lt_{i_0}) \cup \cdots \cup \scp(\lt_{i_{n-1}} | \lt_{i_{n-2}})\big)$ for $n \geqslant 2$. On the other hand, if the construction involves no set partition, then $\alpha = \bigcup \scp(\lt_i)$ for $i = (i_0, i_1, \ldots, i_n)$, where $i_0\hskip-0.03em \in \Li, \,i_{1\!} \in \Li'(\lt_{i_0}), \ldots, i_n\hskip-0.09em \in \Li'(\lt_{i_m} | \lt_{i_l})$, thus $\lt_{i_0\hskip-0.05em} \!\prec\hskip-0.09em \lt_{i_1\hskip-0.09em} \!\prec\hskip-0.09em \cdots \hskip-0.05em\prec\hskip-0.09em \lt_{i_n\!}$. Note that there is no need to construct $\Fm'(\lt_i)$ in \hyperref[alg:Scan]{\texttt{Scan}}/\hyperref[alg:Scope]{\texttt{Scope}} L:9 (cf. Algorithm \ref{alg:CnstAssg}).
\end{note*}

For instance, if Example \ref{e:example3} involves no set partition, then $\alpha = \{\scp(\nx7), \scp(x_2), \scp(x_1)\}$, in which $\scp(\nx7) = \color{purple}\{\nx7, \nx6\}$, $\scp(x_2) = \color{green!40!black}\{x_2\}$, and $\scp(x_1) = \{x_1, \color{green!40!black}x_2 \color{black}, \color{purple}\nx7 \color{black}, \color{purple} \nx6 \color{black}\}$. Also, $\nx7 \hskip-0.07em\prec\hskip-0.07em x_2 \hskip-0.07em\prec\hskip-0.07em x_{1\hskip-0.12em}$ due to $x_2 \hskip-0.07em \in \Fm'(\nx7)$ and $x_1 \hskip-0.13em \in \Fm'(x_2 | \nx7)$. Moreover, $\scp(\nx7)$, $\scp(x_2 | \nx7)$, and $\scp(x_1 | x_2)$ form a partition over the set $\Fm$, where $\scp(x_2 | \nx7) = \scp(x_2) - \scp(\nx7)$ and $\scp(x_1 | x_2) = \scp(x_1) -  \big(\scp(x_2 | \nx7) \cup \scp(\nx7)\big)$. As a result, $\alpha = \Fm(\nx7, x_2, x_1) = \{\nx7, \nx6\} \cup \{x_2\} \cup \{x_1\}$ such that $\{\nx7, \nx6\} \cap \{x_2\} \cap \{x_1\} = \emptyset$.

\end{document}